\documentclass[fleqn,usenatbib]{mnras}

\usepackage{newtxtext,newtxmath}

\usepackage[T1]{fontenc}

\DeclareRobustCommand{\VAN}[3]{#2}
\let\VANthebibliography\thebibliography
\def\thebibliography{\DeclareRobustCommand{\VAN}[3]{##3}\VANthebibliography}

\usepackage{graphicx}	
\usepackage{amsmath}	
\usepackage{soul}
\usepackage{hyperref}
\usepackage{gensymb}

\title[Variations of Ca II K Line Profile Parameters]{Variations of the Ca II K Line Profile Parameters with Solar Latitude and Time Observed from Kodaikanal Solar Observatory}

\author[A. Srinivasa et al.]{
Apoorva Srinivasa,$^{1}$\thanks{E-mail: ampsr4phy22001@am.students.amrita.edu (AS)}
Anu Sreedevi,$^{2}$
K P Raju,$^{3}$
K Nagaraju,$^{3}$
Jagdev Singh,$^{3}$
\newauthor{Narayanankutty Karuppath,$^{1}$
P Devendran,$^{4}$
T Ramesh Kumar$^{4}$
and P Kumaravel$^{4}$}
\\
$^{1}$Department of Physics, Amrita Vishwa Vidyapeetham, Amritapuri, 690525, India\\
$^{2}$Department of Physics, Indian Institute of Technology (Banaras Hindu University), Varanasi, 221005, India\\
$^{3}$Indian Institute of Astrophysics, Koramangala, Bengaluru, 560034, India\\
$^{4}$Indian Institute of Astrophysics, Kodaikanal, 624103, India
}

\date{Accepted XXX. Received YYY; in original form ZZZ}

\pubyear{\the\year{}}

\begin{document}
\label{firstpage}
\pagerange{\pageref{firstpage}--\pageref{lastpage}}
\maketitle

\begin{abstract}
The Calcium-K line profiles as functions of solar latitude and time were obtained through our observations from the Kodaikanal Solar Observatory using the solar tunnel telescope and spectrograph with a CCD detector. Observations were conducted on all days with favourable sky conditions. We analysed the data collected over a period of about ten years to study the variations in the Ca II K line profiles recorded between 2015 and 2024, of which 709 days of data were found useful. The temporal and time-averaged latitudinal variations of the K$_{1}$ width, K$_{2}$ width, K$_{3}$ intensity and the intensity ratios of K$_{2v}$/K$_{2r}$ and K$_{2v}$/K$_{3}$ were computed using a semi-automated program. The parameters showed asymmetric increases towards the higher latitudes, with the rates of increase being higher in the southern hemisphere. The temporal plots for K$_{1}$ width and K$_{3}$ intensity showed positive correlations with the plage and spot filling factors, whereas the temporal plots for K$_{2}$ width, K$_{2v}$/K$_{2r}$ and K$_{2v}$/K$_{3}$ intensity ratios showed negative correlations. The time-averaged latitudinal plot for K$_{1}$ width has small peaks near 25$\degree$N and 20$\degree$S. The K$_{2}$ width has a small peak at 0$\degree$. The K$_{3}$ intensity has peaks at 20$\degree$N and 15$\degree$S. The K$_{2v}$/K$_{2r}$ intensity ratio shows peaks at 50$\degree$N, 0$\degree$ and 40$\degree$S. The K$_{2v}$/K$_{3}$ intensity ratio shows peaks at 60$\degree$N, 0$\degree$ and 60$\degree$S. Slope profiles show spectral response to magnetic activity peaks near K$_{3}$ with north-south asymmetries. Such variations in the line profiles are important in the studies of solar irradiance, surface flux transport and solar dynamo.
\end{abstract}




\section{Introduction} \label{sec:intro}

The variability of the Sun’s magnetic field and radiative output significantly impacts not only solar phenomena but also terrestrial climate and space weather. These variations span a wide range of timescales, from short-term (seconds) to long-term (years) processes \citep{2010RvGeo..48.4001G}. The short-term variations include granulation life (several minutes), sunspot movements (few days) and faculae (few days to weeks). The long-term variations include the solar cycle ($\sim$11 years) and solar irradiance. Several indices and parameters -- solar radio flux, solar irradiance, ultraviolet index, Ca-K emission index and geomagnetic indices -- are used to understand and quantify solar variability \citep{2015LRSP...12....4H}. \par
About 63\% of the variation in the Total Solar Irradiance (TSI) occurs in the UV part of the solar spectrum \citep{2011SoPh..269..253M}. Although only 1\% of electromagnetic radiation comes from the Sun at wavelengths smaller than 300 nm, it may contribute more than that to the variations in total irradiance. The fluctuation resulting from large sunspot regions offsets the change in solar irradiance induced by the bright faculae (\citealt{1991suti.conf...11F}; \citealt{1997ESASP.415..251P}). The understanding of how different features, including plages, networks, sunspots and others, contribute to irradiance variability has advanced significantly in the last few years \citep{2013ARA&A..51..311S}. Over the course of the 11-year solar cycle, the Ca-K index fluctuates by 18\% \citep{1981ApJ...249..798W}. The fluctuation in total irradiance caused by a solar cycle has an amplitude of 0.1\% \citep{1998GeoRL..25.4377F}. The correlation between the magnetic field of the sun and Ca-K emission makes spectroscopic investigation of the solar chromosphere important (\citealt{1998SoPh..179..253N}; \citealt{2005MmSAI..76.1018O}). \par
The solar magnetic field, generated by the solar dynamo, drives much of the sun’s activity. Understanding this magnetic field's evolution is essential to studying solar activity cycles and their impact on solar irradiance. A key component of the solar dynamo is the migration of the sunspot regions to the solar equator (\citealt{2020LRSP...17....4C}; \citealt{2000ARep...44..759M}). The variation in Ca-K line profiles across latitudes suggests that magnetic flux transfer takes place on the surface of the sun \citep{2014ApJ...792...22S}. It may be noted that the canopy effect and geometrical foreshortening affect the measurements made above 75$\degree$ latitude, causing poor approximation of the flux distribution \citep{2000SoPh..195..247W}. The variations in Ca-K line profiles are mainly caused by plages and chromospheric networks \citep{2024MNRAS.527.2940C}. The temporal variation of the strength of the underlying magnetic field primarily contributes to the differences in intensities of distinct Ca-K plages, which may also vary throughout different phases of the solar cycle \citep{2015JApA...36...81S}. \par
\begin{figure}
	\centering
	\setkeys{Gin}{draft=False}
	\includegraphics[width=\hsize]{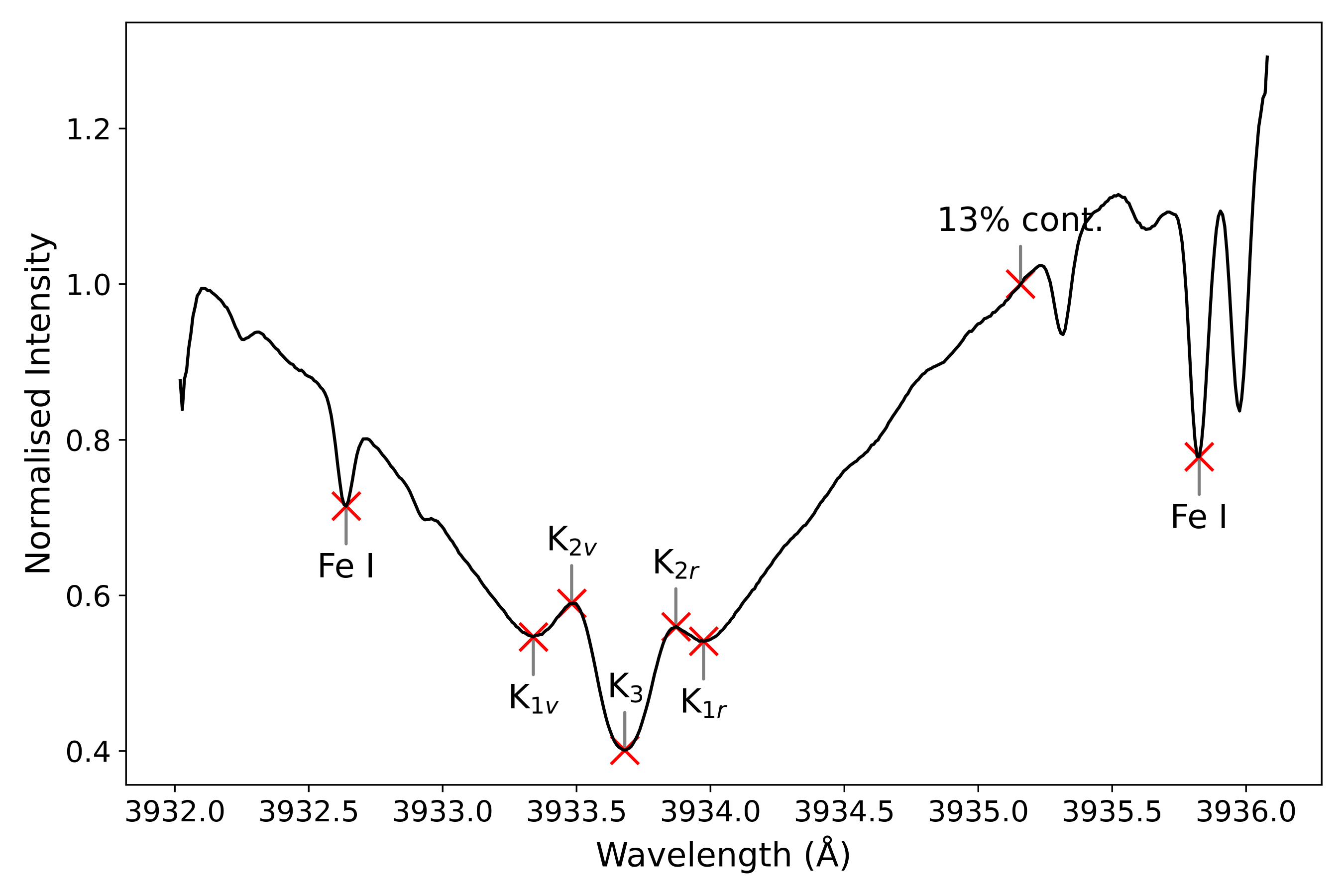}
	\caption{A typical Ca-K line profile from our observations. The three principal features (K$_{1}$, K$_{2}$ and K$_{3}$) of the profiles are marked.}
	\label{cakprof}
\end{figure}

The calcium line was first detected by Joseph Fraunhofer in 1814, and it was included as K in his list of prominent absorption features. \citet{1904ApJ....19...41H} introduced designations for three parts of the Ca-K line profiles. These are shown in Figure \ref{cakprof}, which is an example of a typical Ca-K line profile from our observations. In the figure, the central absorption is called K$_{3}$. The emission features on either side of K$_{3}$ are K$_{2v}$ and K$_{2r}$, and the minima in the profile wings beyond the emission peaks are K$_{1v}$ and K$_{1r}$. The wavelength difference between K$_{2v}$ and K$_{2r}$ is the K$_{2}$ width, and the difference between K$_{1v}$ and K$_{1r}$ is the K$_{1}$ width. The source function -- the ratio of the emission coefficient to the absorption coefficient -- for the K line reflects the value of the local temperature of the region \citep{1981ApJS...45..635V}. The increase in temperature results in an increase in source function. At about 1900 kilometers, higher temperatures result in an increased population of ionised calcium atoms. These ionised calcium atoms effectively absorb radiation at the specific wavelength corresponding to the calcium-K line, leading to the formation of the dark core reversal (K$_{3}$) $\approx$1700--2000 km from the solar surface. At lower heights, the source function is affected by scattering due to lower temperatures, thus contributing to the formation of wings (K$_{1}$ minima) $\approx$400--600 km. This mechanism causes the existence of two bumps in the line profile near its core (K$_{2}$ peaks), at $\approx$700--1500 km. The Ca-K line is particularly interesting to solar astronomers studying solar variability because it provides information about the activity and dynamics of the sun's chromosphere \citep{1987ApJ...313..456S}. \par
It was observed that the Ca-K line profiles change with the solar cycle (\citealt{1983A&A...124...43O}; \citealt{1978ApJ...226..679W}; \citealt{2020ApJ...897..181B}). From the spatially resolved spectra, it was found that the areas of plages and networks exhibit larger K$_{1}$ widths and smaller K$_{2}$ widths when compared to those of quiet regions (\citealt{1981ApJ...249..798W}; \citealt{1971SoPh...17..316B}). In general, it was found that K$_{2v}$ has a larger intensity than K$_{2r}$. The lifespan of sunspots influences the K$_{1}$ and K$_{2}$ widths \citep{2022FrASS...938949C}, which in turn reflect the extent to which the area and intensity of features vary every day. These variations in K$_{1}$ and K$_{2}$ widths, as well as the intensities at the wavelengths of these emission peaks and wings, provide valuable insights into the magnetic activity occurring in solar plages and networks, important features that influence the solar cycle's impact on solar irradiance \citep{2015JApA...36...81S}. In order to study limb darkening, K$_{2}$ widths as a function of longitude have been analysed by \citet{1966ApNr...10..101E}. \par
Since 1969, the sun-as-a-star observations in the chromospheric Ca-K line have been conducted at the Kodaikanal Solar Observatory (lat. 10$\degree$14' N; long. 77$\degree$28' E). From 1969 to 1984, the changes in Ca-K line parameters with the solar cycle were studied by using the solar disk-averaged Ca-K line profiles \citep{1987ApJ...313..456S}. Since 1986, chromospheric variations with both the solar cycle and latitude have been analysed using Ca-K spectra acquired as a function of latitude and integrated over longitudes by \citet{1988KodOB...9..159S}. By isolating the contributions of plages and networks, the study was able to more accurately assess the variability in the Ca-K line width. This methodology of integrating spectra over visible longitudes at specific latitude belts was developed primarily to study the temporal variations of chromospheric activity as a function of solar latitude over extended periods, covering multiple solar cycles. While local line formation physics depends on parameters such as surface element velocity and centre-to-limb angle (\citealt{2024MNRAS.527.2940C}; \citealt{2017SSRv..210...37L}), the latitude-resolved approach offers unique advantages for tracking large-scale solar dynamics. The relationships between the different coordinate variables (line-of-sight velocity, centre-to-limb angle, and latitude) are presented in Appendix \ref{sec:app_a}.\par
Figure \ref{composites_plot} shows composite images from the Solar Dynamics Observatory's (SDO) Atmospheric Imaging Assembly (AIA) at 304 {\AA} and 1600 {\AA}. We constructed these composites from over 100 individual AIA images (from 2023 and 2024) acquired during our observation period using the max stacking method. This method records the maximum value at each pixel from all the images in the pool and generates the composite image. As can be seen from the composite images, the activity regions are concentrated between 40$\degree$N and 40$\degree$S. A Butterfly-diagram-like pattern is visible in the composites of both 304 {\AA} and 1600 {\AA}. \par
\begin{figure}
	\centering
	\setkeys{Gin}{draft=False}
	\includegraphics[width=\hsize]{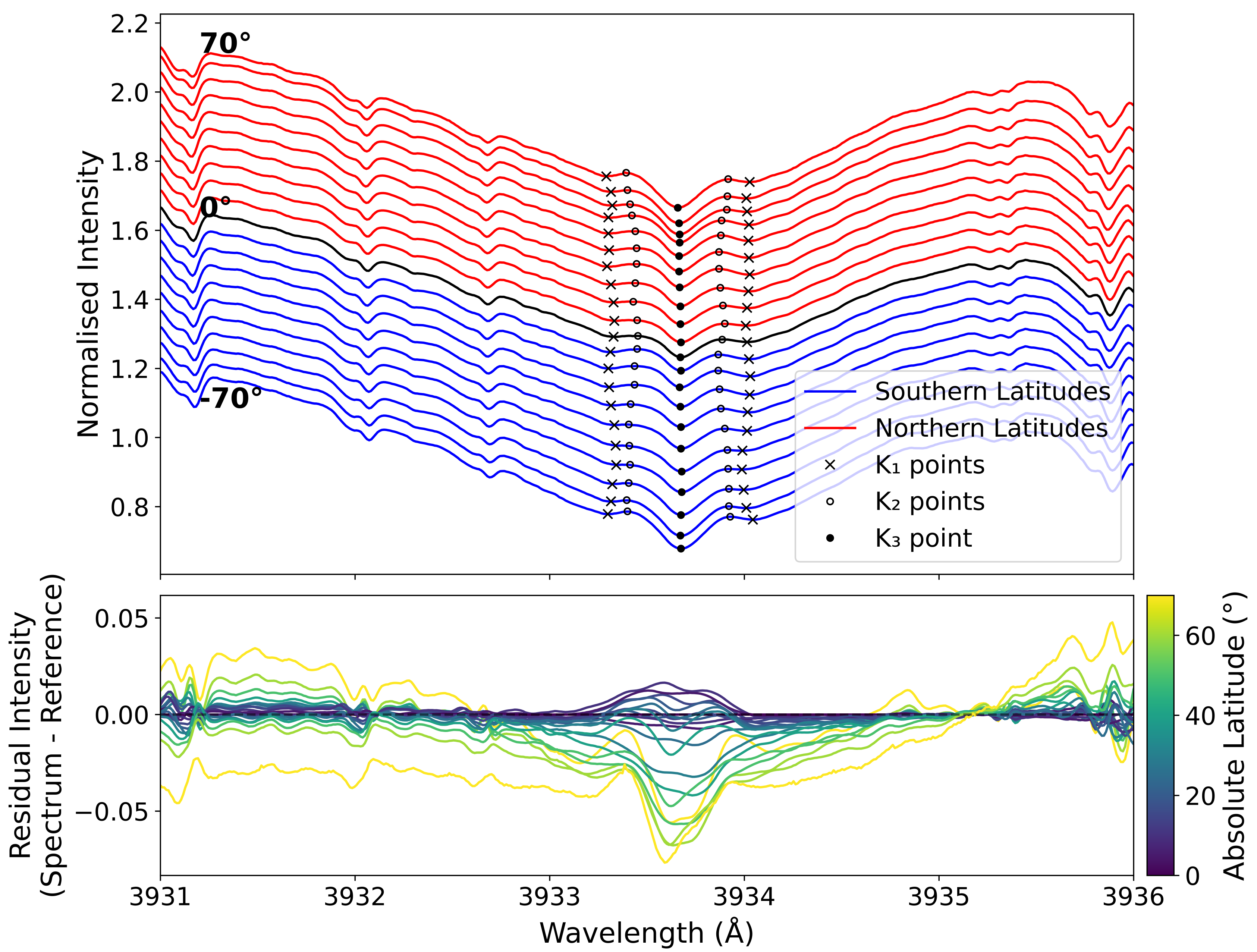}
	\caption{(Top) Latitude-wise Ca-K line profiles from our observations. Starting from -70$\degree$, the profiles are successively offset vertically by 0.05 units for visual clarity. (Bottom) Residual intensity plot of the profiles taking the profile at the equator as the reference.}
	\label{cakproflats}
\end{figure}

The latitude-wise analysis with longitude integration provides valuable insights into the distribution and evolution of magnetic activity across solar latitudes over time. While our Ca-K line parameters are intensity metrics rather than direct velocity measurements, they serve as effective tracers of magnetic features (plages, enhanced network) that are passively transported by large-scale plasma flows. Meridional circulation, characterised by polewards flows at the surface and equatorwards return flows in the interior \citep{2023ApJ...944..218P}, is known to transport magnetic flux as part of the solar dynamo mechanism. By analysing the systematic latitudinal migration of magnetic activity through our Ca-K intensity proxies, we can identify patterns consistent with the transport effects of meridional flows. This approach allows us to observe the consequential signatures of these flows rather than measuring the flows themselves, providing indirect evidence of their influence on magnetic feature distribution across different latitudes in both hemispheres. Such observations contribute to our understanding of the transport mechanisms involved in solar dynamo processes \citep{2014ApJ...792...22S}. Extensive studies have been conducted on the fluctuations of Ca-K line profiles as functions of latitude and time during solar cycles 22 and 23 by \citet{2015IIA...PhD...Thesis}. The study included in-depth analyses of the meridional flow patterns using the Ca-K line profiles, providing valuable insights into the mechanism of the solar dynamo. Figure \ref{cakproflats} shows the Ca-K line profiles plotted for several latitudes to illustrate the variations across latitudes. The figure also includes a residual plot made using the equator profile as the reference. It shows that the intensities at higher latitudes are lower than those of lower latitudes at the core of the profiles. \par
While our method is not directly designed to study differential rotation (where different latitudes rotate at different speeds), the latitude-dependent evolution of chromospheric features tracked through Ca-K emission can indirectly reflect aspects of differential rotation (\citealt{2020ApJ...897..181B}; \citealt{1984ApJ...276..766K}). Similarly, though we do not resolve individual granules or supergranules, the cumulative effects of granular and supergranular convection and diffusion influence the transport and distribution of magnetic flux observed in the chromospheric network through Ca-K emission (\citealt{2014SSRv..186..251N}; \citealt{2000SoPh..195..247W}). By obtaining a statistical average of chromospheric conditions at different latitude belts through longitude integration, we reduce the impact of localised, short-lived features and focus on longer-term, latitude-dependent trends relevant to the solar dynamo. \par
\begin{figure}
	\centering
	\setkeys{Gin}{draft=False}
	\includegraphics[width=\hsize]{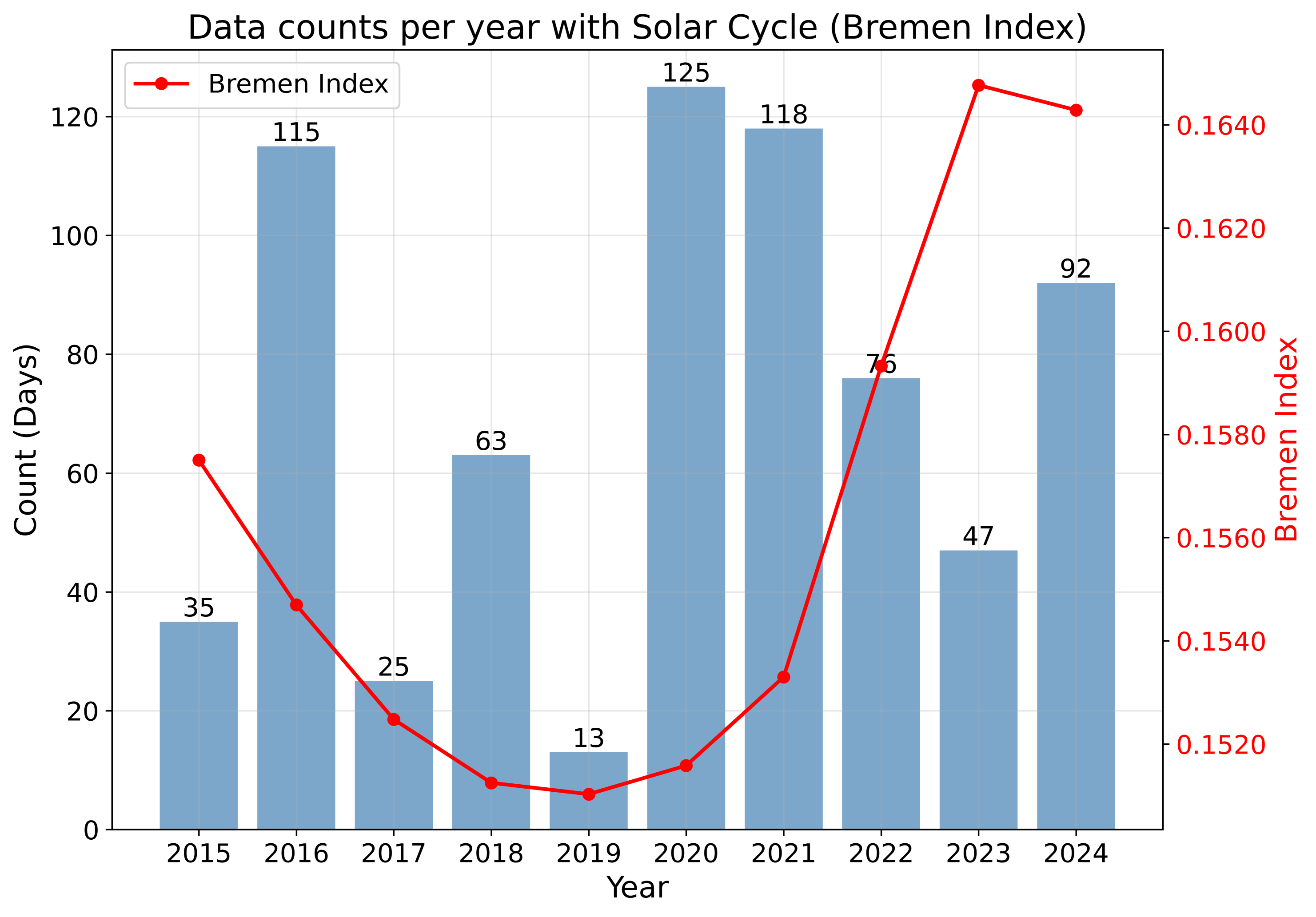}
	\caption{Barchart showing the spread of our data over 2015 to 2024. The solar irradiance variability over the years is shown by the Bremen composite Mg II index or Bremen Index.}
	\label{yearly_data_counts}
\end{figure}

Our study extends the analysis of chromospheric variations by examining solar cycle-dependent changes in the Ca-K line profiles obtained for a period of about ten years, from 2015 to 2024, and contains 709 days of useful data (Figure \ref{yearly_data_counts}) to examine the variation of Ca-K line parameters as functions of latitude and time. In section \ref{sec:obsanalysis}, we provide the details of the observations from the Kodaikanal Solar Observatory (KSO) and the methodology used for getting the data ready for analysis, followed by a list of parameters used for data analysis. Sections \ref{sec:results}, \ref{sec:discussion} and \ref{sec:conclusion} detail the results obtained, the discussions on those and conclusions.

\begin{figure}
	\centering
	\setkeys{Gin}{draft=False}
	\includegraphics[width=\hsize]{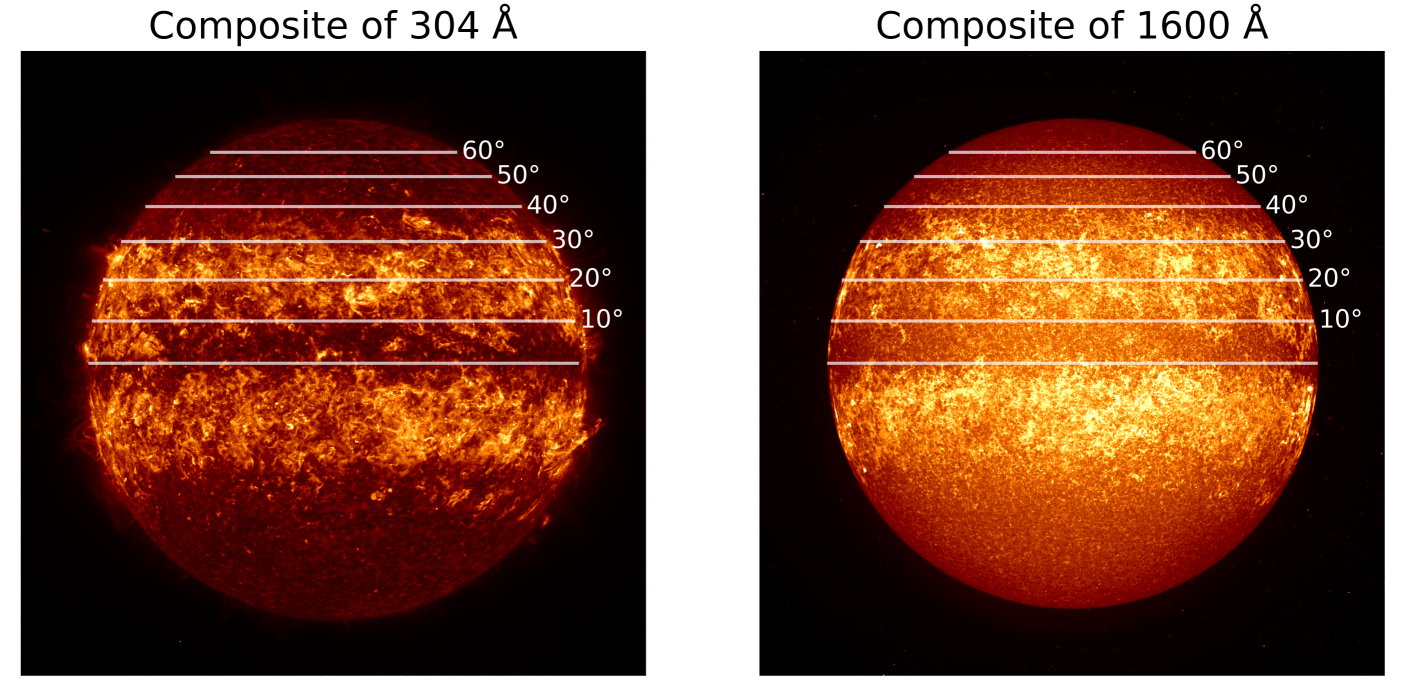}
	\caption{Composite images of the sun at (Left) 304 {\AA} and (Right) 1600 {\AA} constructed from the data overlapping with our observation days. (Original data are courtesy of SDO's AIA)}
	\label{composites_plot}
\end{figure}

\section{Observations and Data Analysis} \label{sec:obsanalysis}

\subsection{Observations} \label{subsec:obs}
The longitudinally integrated Ca-K spectra have been recorded at KSO as a function of solar latitude since 1986 \citep{1988KodOB...9..159S}. The spectra were recorded using a CCD camera of 2K $\times$ 2K format with a pixel size of 13.5 $\times$ 13.5 microns. The image scale on this detector is 0.07"/pixel, the spectral sampling is 5.6 m{\AA}/pixel, and the spectral resolution (limited by the 100 $\mu$m slit width) is approximately 11 m{\AA}, giving an equivalent R of $\approx$350,000. The observations are carried out during the morning hours due to the better-seeing conditions and require a cloud-free sky for about 45 minutes. This is to minimise atmospheric turbulence and ensure optimal image quality during acquisition. The observing days are mostly confined to the months between September and March and occasionally a few days from April to August as the other days are affected by the southwest and northeast monsoons in Kodaikanal. \par
We have adopted \citet{2015IIA...PhD...Thesis}'s imaging method of the sun's spectra -- improved upon \citep{1988KodOB...9..159S} -- around the Ca-K line wavelength as a function of latitude integrated across longitude as follows:
\begin{enumerate}
	\item Sun charts are made considering the size of the sun's image and proper ephemerides (such as the inclination of the sun or B$_{0}$ angle) such that the centre of the chart is located at the B$_{0}$ angle.
	\item The sun chart is set up such that its marked north-south axis is parallel to the sun's image's rotation axis.
	\item A fraction of the sun’s image is allowed to pass through the spectrograph’s slit in such a way that when the North limb of the sun's image moves along a latitude line, the same latitude falls on the slit centre.
	\item To get the Ca-K wavelength spectra, the sun's image is moved uniformly from its east to west end on the spectrograph's slit at specific latitudes.
	\item The widths of the latitude belts taken are 10$\degree$ in ±(30$\degree$ -- 70$\degree$) and 5$\degree$ in (0$\degree$ -- ±30$\degree$).
\end{enumerate}

\subsection{Calculation of Profile Parameters} \label{subsec:profparam}
The data were corrected for dark and flat field variations. We developed a semi-automated program to perform the following:
\begin{enumerate}
	\item A slice from any spectral image shows us the line spectrum. In order to deal with some of the noise or artifacts that may have slipped the corrections, from each data file, we considered around 300-350 slices to obtain the averaged spectra for that particular latitude. 
	\item For the wavelength calibration, the photospheric absorption lines of Fe I (shown in \ref{cakprof}) at 3932.640 {\AA} and 3935.825 {\AA} were used. Rough windows were specified around these points and fitting algorithms were used within the windows to extract the exact positions of these points automatically.
	\item To normalise the widely varying intensity values, we used the residual intensity value (13\% of the continuum as determined by \citet{1968SoPh....3..523W}) at 3935.160 {\AA} (shown in \ref{cakprof}) to scale the intensity values in each spectrum.
	\item We extracted most of the data points automatically using fitting algorithms. Those data points that had values that went beyond the tolerance set in the algorithms were extracted by locating the points manually.
	\item Lastly, we removed the outliers by considering the 2$\sigma$-window for each year.
\end{enumerate}
The parameters of the Ca-K line profiles analysed in this study are K$_{1}$ width, K$_{2}$ width, K$_{3}$ intensity, K$_{2v}$/K$_{2r}$ intensity ratio and K$_{2v}$/K$_{3}$ intensity ratio.

\subsection{Calculation of Spot and Plage Filling Factors} \label{subsec:fillingfactors}
To investigate the relationship between Ca-K line profile parameters and solar activity features, we calculated the spot and plage filling factors (FF) using the SolAster package \citep{2022ascl.soft07009E} for the days corresponding to our Ca-K observations. We utilised the Helioseismic and Magnetic Imager (HMI) data from the SDO for the period from 2015 to 2024, which covers the majority of our observational period. For each day of our Ca-K observations, the corresponding SDO/HMI images were retrieved and processed to calculate the sun-as-a-star (or disk-integrated (DI)) plage + spot FF as well as latitude-wise (LW) FF. Since the spot FF tend towards 0 beyond ±40$\degree$ latitudes, where minimal spot activity is observed, we decided to consider (plage + spot) FF for our analyses.\par
The relationship between the calculated FF and the Ca-K line profile parameters was then assessed through correlation analysis to identify possible connections between photospheric/chromospheric features and the spectral characteristics observed in our study. This analysis complements the traditional sunspot number correlation and provides additional insights into how different magnetic features contribute to the observed spectral variations (\citealt{1987ApJ...313..456S}; \citealt{2024MNRAS.527.2940C}).

\subsection{Calculation of Flux Difference per Filling Factor Difference} \label{subsec:fluxffdiff}
To investigate the activity profile per unit of filling factor at the mean $\mu$ angle, we did the following:
\begin{enumerate}
	\item Extracted the quiet sun latitude profiles for each of the latitudes by considering only those days in 2020 when zero sunspots were reported. These profiles were averaged latitude-wise to obtain the reference profiles.
	\item From each day's profiles from our time period, we differenced the reference profiles latitude-wise to obtain the "corrected" profiles.
	\item Next, we performed similar steps with FF data, again, latitude-wise and considered (plage + spot) FF.
	\item With these data in hand, we considered a 1.2 {\AA} window around 3933.6 {\AA} to calculate the flux difference per filling factor difference at each wavelength point in the selected window.
	\item A linear fit was performed between the time series of flux differences and the time series of filling factor differences.
	\item The slope coefficients obtained were plotted against the corresponding wavelengths at which they were obtained.
\end{enumerate}

\subsection{Sources of Uncertainty} \label{subsec:uncertainty}
Observational data inherently contain uncertainties. The possible sources of uncertainty in our data include:
\begin{itemize}
	\item Environmental factors such as atmospheric distortions, air mass and weather variability may have affected the data collected.
	\item CCD degradation over time due to pixel defects, declining charge transfer efficiency, etc. may have resulted in small errors.
	\item The correction data collected for dark currents and flat fields may not have removed the noise and artefacts completely.
	\item Since the data extraction was done using fitting, there may have been minor inaccuracies due to imperfect fits.
	\item Manual location and extraction of data points may have contributed to errors due to the presence of human elements.
\end{itemize}

\section{Results} \label{sec:results}
After computing the above-mentioned parameters from the data for each day of observations, we describe their variations as functions of time and solar latitude in the following subsections. Temporal variation plots are shown along with the FF data for the selected time period between August 2015 and April 2024. The time-averaged latitude-wise measurements -- their mean and standard deviation ranges -- for each parameter are summarised in Table \ref{tab1}. \par

\begin{table*}
	\centering
	\caption{Mean and standard deviations for the latitude-wise variations of the parameters.}
	\label{tab1}
	\begin{tabular}{|c|c|c|c|c|c|}
		\hline \multicolumn{6}{|c|}{Quantified Variations of Various Ca-K Line Parameters} \\ \hline
		\hline{\textbf{Latitude ($\degree$)}} & {\textbf{K$_{1}$ Width ({\AA})}} & {\textbf{K$_{2}$ Width ({\AA})}} & \textbf{K$_{3}$ Intensity} & \textbf{K$_{2v}$/K$_{2r}$} & \textbf{K$_{2v}$/K$_{3}$} \\ & & & \textbf{(Normalised)} & \textbf{Intensity Ratio} & \textbf{Intensity Ratio} \\ \hline
	-70 & 0.739 ± 0.042 & 0.462 ± 0.030 & 0.062 ± 0.012 & 1.048 ± 0.021 & 1.425 ± 0.126 \\
	-60 & 0.690 ± 0.036 & 0.431 ± 0.022 & 0.061 ± 0.012 & 1.056 ± 0.022 & 1.426 ± 0.119 \\
	-50 & 0.650 ± 0.029 & 0.409 ± 0.015 & 0.060 ± 0.012 & 1.061 ± 0.022 & 1.412 ± 0.122 \\
	-40 & 0.625 ± 0.029 & 0.397 ± 0.014 & 0.061 ± 0.012 & 1.062 ± 0.023 & 1.387 ± 0.114 \\
	-30 & 0.617 ± 0.037 & 0.389 ± 0.016 & 0.063 ± 0.012 & 1.061 ± 0.023 & 1.362 ± 0.108 \\
	-25 & 0.624 ± 0.049 & 0.382 ± 0.020 & 0.065 ± 0.013 & 1.058 ± 0.023 & 1.337 ± 0.105 \\
	-20 & 0.627 ± 0.055 & 0.377 ± 0.019 & 0.066 ± 0.013 & 1.058 ± 0.024 & 1.326 ± 0.102 \\
	-15 & 0.622 ± 0.057 & 0.375 ± 0.022 & 0.066 ± 0.014 & 1.058 ± 0.023 & 1.320 ± 0.101 \\
	-10 & 0.613 ± 0.054 & 0.374 ± 0.021 & 0.066 ± 0.014 & 1.059 ± 0.024 & 1.322 ± 0.102 \\
	-5 & 0.600 ± 0.048 & 0.377 ± 0.022 & 0.065 ± 0.014 & 1.060 ± 0.023 & 1.322 ± 0.105 \\
	0 & 0.591 ± 0.037 & 0.379 ± 0.024 & 0.065 ± 0.014 & 1.061 ± 0.023 & 1.323 ± 0.105 \\
	5 & 0.595 ± 0.039 & 0.377 ± 0.022 & 0.065 ± 0.013 & 1.060 ± 0.023 & 1.320 ± 0.103 \\
	10 & 0.606 ± 0.047 & 0.377 ± 0.023 & 0.066 ± 0.013 & 1.059 ± 0.023 & 1.317 ± 0.101 \\
	15 & 0.618 ± 0.054 & 0.375 ± 0.022 & 0.066 ± 0.013 & 1.057 ± 0.023 & 1.317 ± 0.100 \\
	20 & 0.627 ± 0.060 & 0.376 ± 0.022 & 0.067 ± 0.013 & 1.056 ± 0.022 & 1.318 ± 0.098 \\
	25 & 0.629 ± 0.059 & 0.379 ± 0.019 & 0.066 ± 0.013 & 1.056 ± 0.023 & 1.326 ± 0.101 \\
	30 & 0.624 ± 0.054 & 0.385 ± 0.019 & 0.065 ± 0.013 & 1.057 ± 0.023 & 1.334 ± 0.104 \\
	40 & 0.622 ± 0.046 & 0.390 ± 0.017 & 0.064 ± 0.013 & 1.057 ± 0.023 & 1.348 ± 0.109 \\
	50 & 0.629 ± 0.031 & 0.399 ± 0.015 & 0.062 ± 0.012 & 1.058 ± 0.021 & 1.373 ± 0.111 \\
	60 & 0.657 ± 0.029 & 0.415 ± 0.017 & 0.062 ± 0.012 & 1.055 ± 0.020 & 1.386 ± 0.115 \\
	70 & 0.704 ± 0.038 & 0.440 ± 0.024 & 0.065 ± 0.012 & 1.044 ± 0.021 & 1.367 ± 0.118 \\ \hline
	\end{tabular}
\end{table*}

\subsection{Variations in K$_{1}$ Widths} \label{subsec:k1width}
\begin{figure}
	\centering
	\setkeys{Gin}{draft=False}
	\includegraphics[width=\hsize]{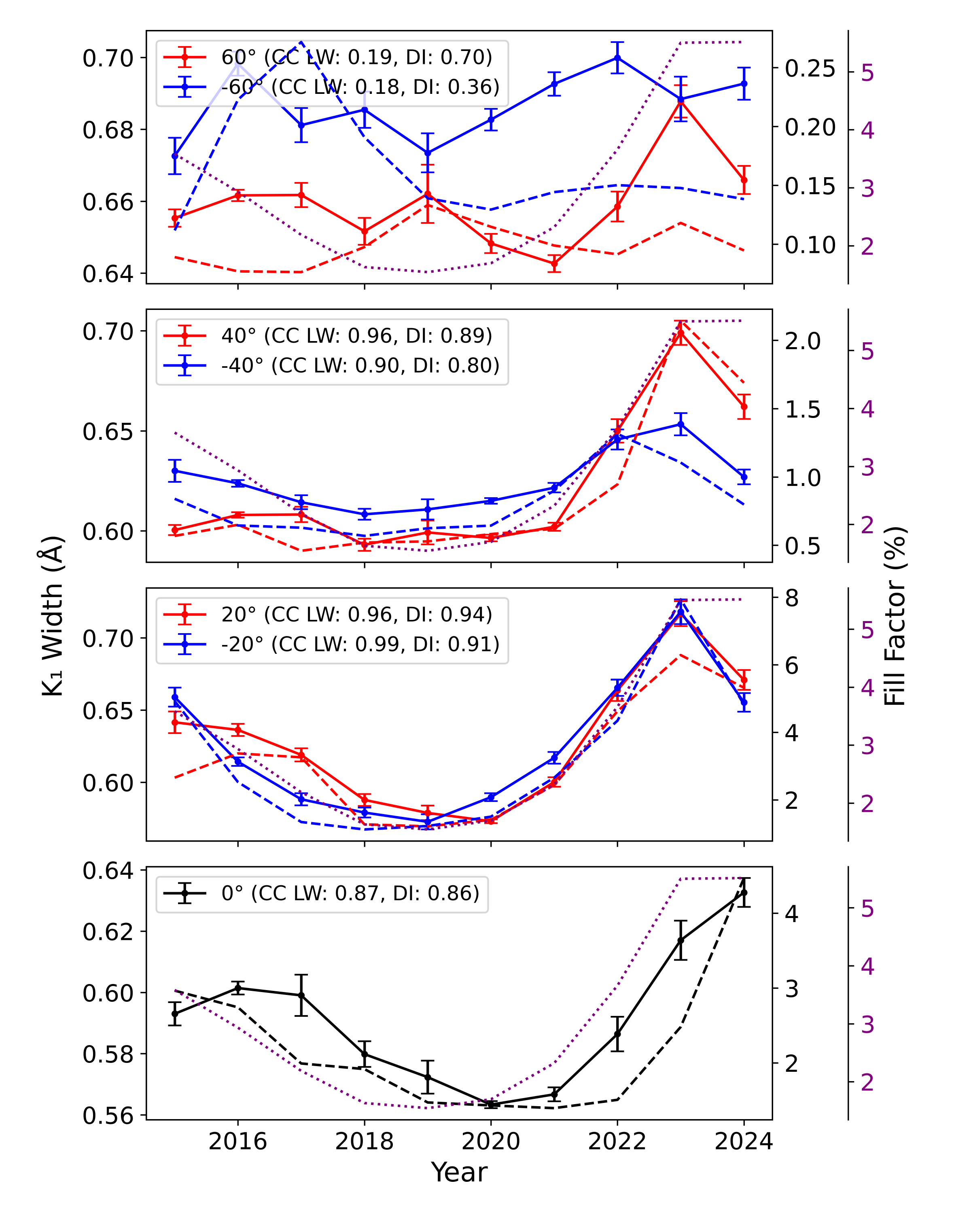}
	\caption{Temporal Variations in K$_{1}$ widths for select latitudes. Circular points with error bars show the yearly-averaged widths, red for northern latitudes and blue for southern. The red and blue (black for 0$\degree$) dashed lines show the yearly-averaged latitude-wise (LW) plage+spot FF for the respective hemispheres. The purple dotted line shows the yearly-averaged disk-integrated (DI) FF.}
	\label{temp_var_k1}
\end{figure}

\begin{figure}
	\centering
	\setkeys{Gin}{draft=False}
	\includegraphics[width=\hsize]{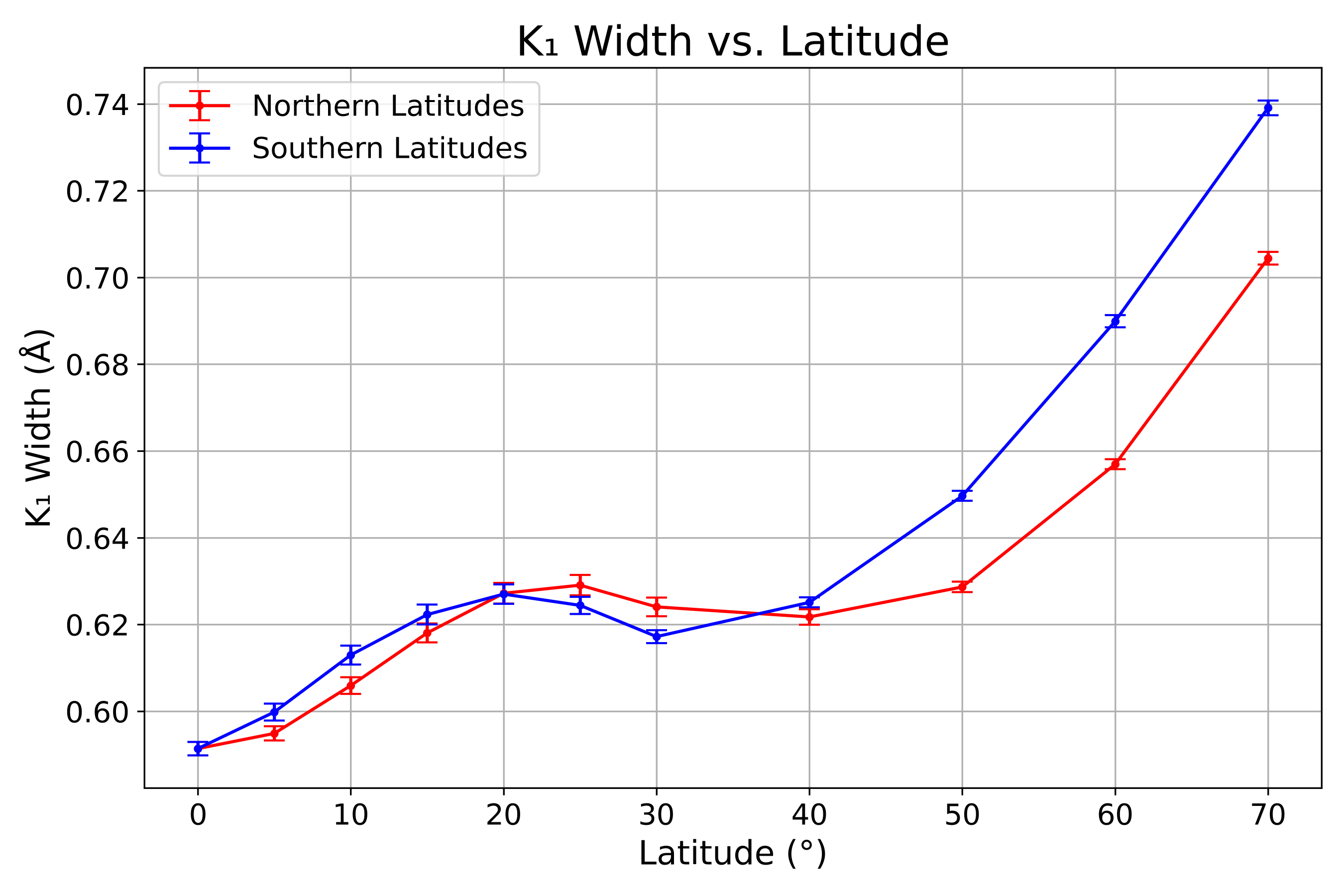}
	\caption{Time-averaged K$_{1}$ width variation as a function of latitude. Northern hemisphere in red and southern hemisphere in blue.}
	\label{k1var}
\end{figure}

Figure \ref{temp_var_k1} shows the variations in the K$_{1}$ widths over time for select latitudes. Each of the latitudes plotted (except 60$\degree$S) shows a similar underlying trend, which is a decrease in the widths during 2015 -- 2019/2020, followed by an increase towards 2024. The trend for 60$\degree$S shows only an increase during 2015 -- 2024. The Pearson Correlation Coefficients (CC) between the K$_{1}$ width and LW-FF as well as DI-FF are positive. The maximum LW- and DI-FF correlations of 0.99 and 0.94 are obtained at the 20$\degree$S and 20$\degree$N latitudes, and minimum correlations of 0.18 and 0.36, both are obtained at 60$\degree$S. \par
Figure \ref{k1var} shows the plot of K$_{1}$ width, averaged over time, as a function of latitude. The width varies in the range of 0.591 -- 0.739 {\AA}, with the standard deviations ranging from 0.029 to 0.060 {\AA}. The figure shows that the K$_{1}$ width generally increases with the latitude in both hemispheres. The K$_{1}$ width shows larger values around 20--25$\degree$N and 20$\degree$S compared to the values for the previous neighbouring latitudes. The widths are generally larger in the southern hemisphere, and their increase is steeper in the southern hemisphere, moving from 40$\degree$S to 70$\degree$S. The averaged K$_{1}$ width over all latitudes varies by 34.42\% between the minimum and maximum phases. \par

\begin{figure}
	\centering
	\setkeys{Gin}{draft=False}
	\includegraphics[width=\hsize]{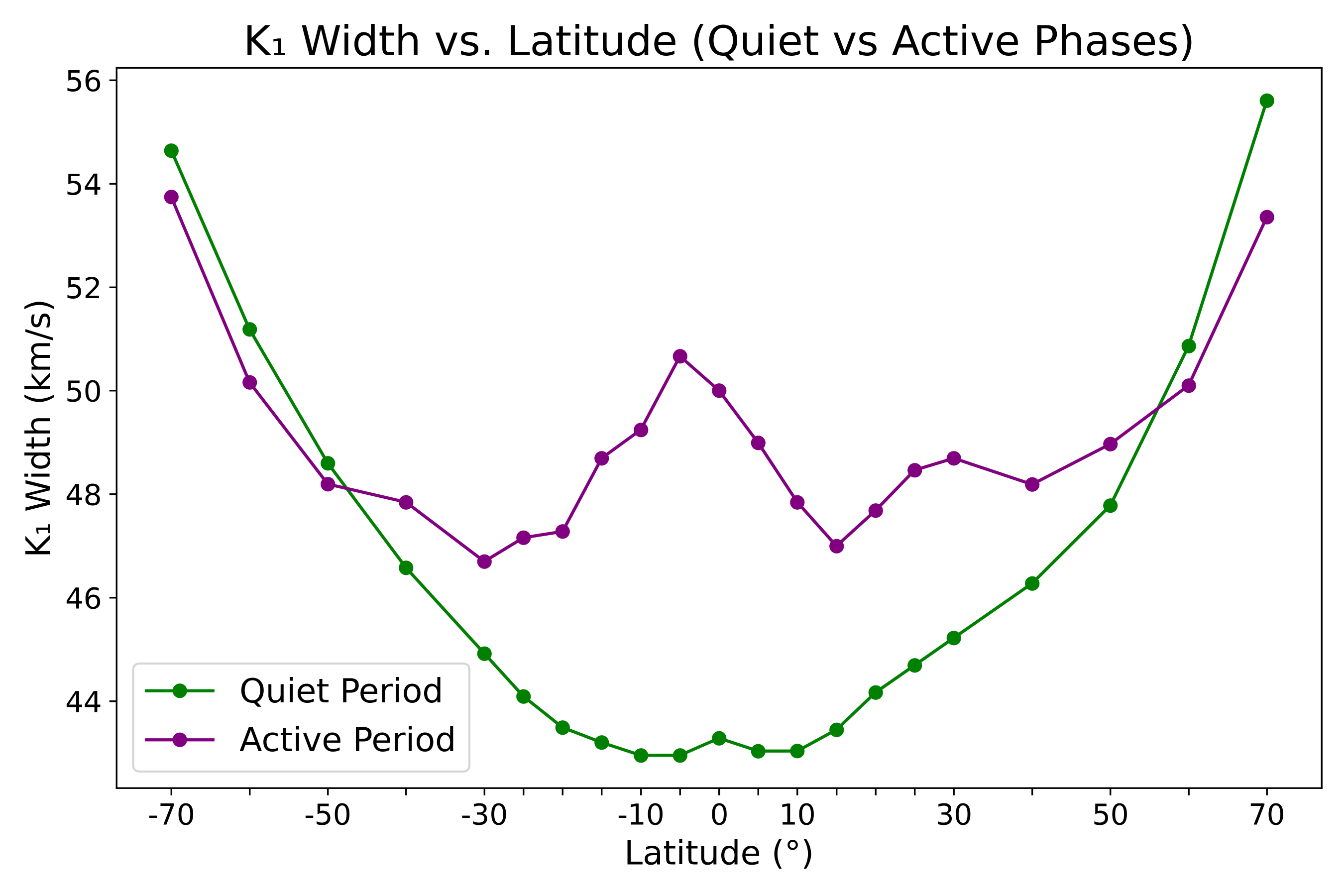}
	\caption{K$_{1}$ width variations as a function of latitude during quiet (green) and active (purple) phases of the sun.}
	\label{k1width_quiet_active}
\end{figure}

Due to the rotational shift of the sun, our solar spectra should be shifted by ±2 km/s between east and west which should introduce a broadening of the line closer to the equator when we integrate the spectra. But we notice from our Table \ref{tab1} that this is not the case, rather the opposite. While there exists broadening due to shifting post-integration, we surmise that the effects of such shifts are quite low due to the magnitudes of the velocities of shifting, as compared to activities that take place. To visualise this, we have plotted a latitude-wise comparison plot between the active and quiet phases in Figure \ref{k1width_quiet_active}. We have chosen and averaged about 7-10 days each in the months of February 2020 and 2024 where the number of sunspots (proxy for activities) were all 0 and greater than 100, respectively.

\subsection{Variations in K$_{2}$ Widths} \label{subsec:k2width}
\begin{figure}
	\centering
	\setkeys{Gin}{draft=False}
	\includegraphics[width=\hsize]{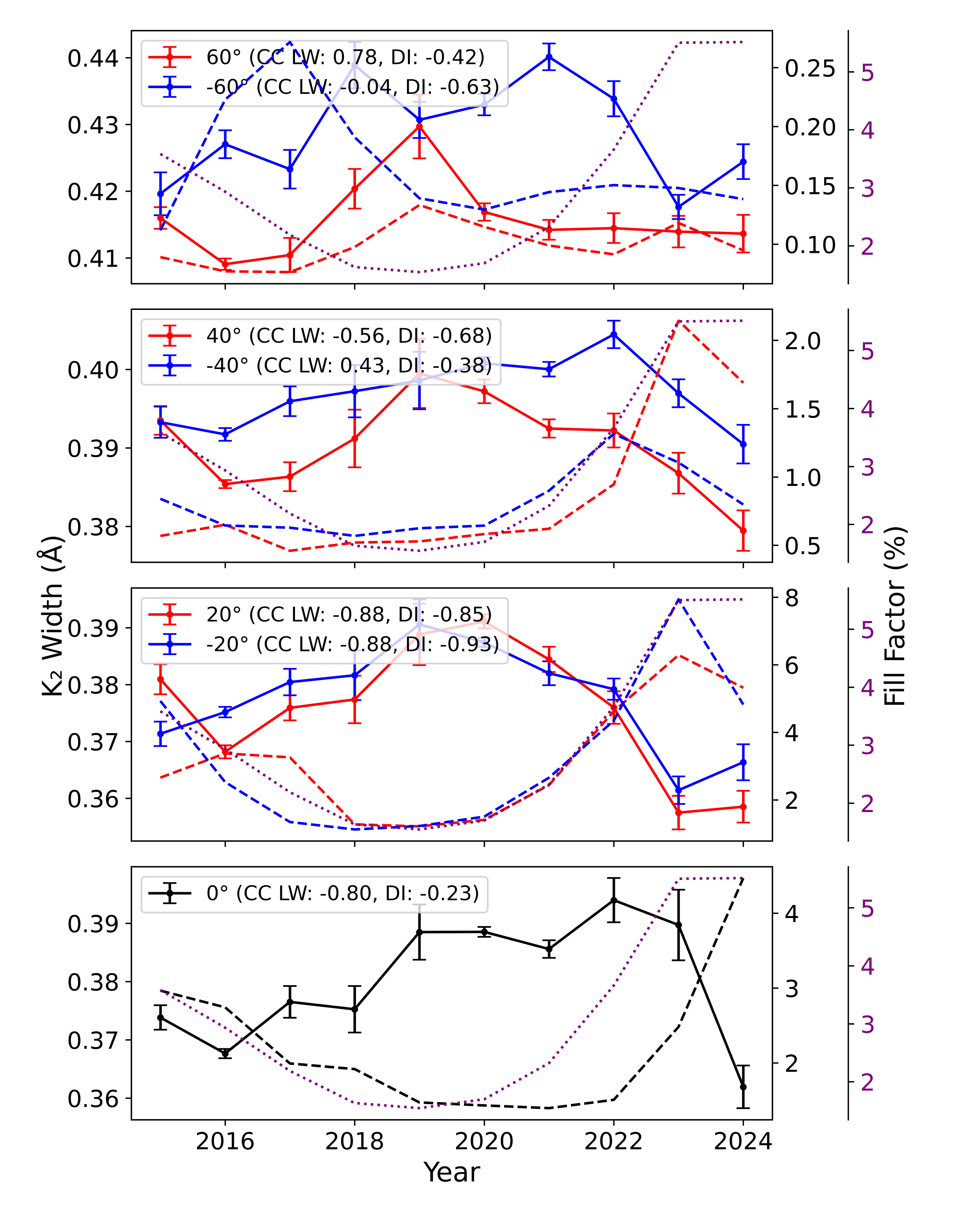}
	\caption{Temporal Variations in K$_{2}$ widths for select latitudes. Circular points with error bars show the yearly-averaged widths, red for northern latitudes and blue for southern. The red and blue (black for 0$\degree$) dashed lines show the yearly-averaged LW plage+spot FF for the respective hemispheres. The purple dotted line shows the yearly-averaged DI-FF.}
	\label{temp_var_k2}
\end{figure}

\begin{figure}
	\centering
	\setkeys{Gin}{draft=False}
	\includegraphics[width=\hsize]{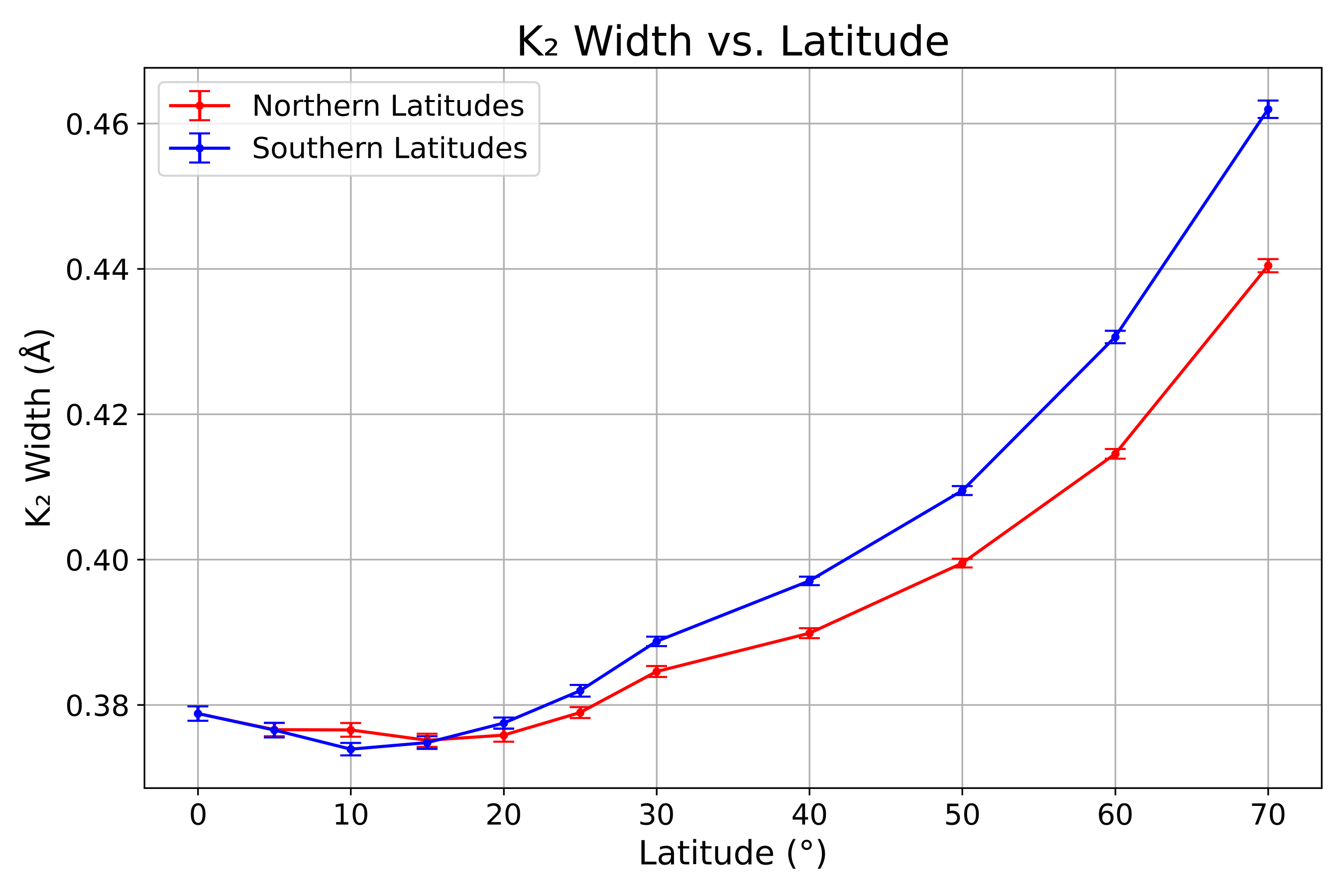}
	\caption{Time-averaged K$_{2}$ width variation as a function of latitude. Northern hemisphere in red and southern hemisphere in blue.}
	\label{k2var}
\end{figure}

Figure \ref{temp_var_k2} shows the variations in the K$_{2}$ widths for select latitudes over time. Most of the latitudes plotted show a similar underlying trend, which is an increase in the widths during 2015 -- 2019/2020, followed by a decrease towards 2024, except latitudes 0$\degree$ and -40$\degree$, which show an increase in the widths until 2022 before decreasing. The CCs between the K$_{2}$ width and LW-FF as well as DI-FF are negative. The maximum LW- and DI-FF negative correlations of -0.88 and -0.93, both are obtained at 20$\degree$S (also at 20$\degree$N for LW-FF), and minimum negative correlations of -0.04 and -0.23 are obtained at 60$\degree$S and 0$\degree$ latitudes. It may be noted that despite the trend line showing a clear negative correlation, the smaller LW-FF CC at 60$\degree$S is due to a few consecutive individual data points that show a positive correlation when considered two at a time. \par
Figure \ref{k2var} shows the plot of K$_{2}$ width, averaged over time, as a function of latitude. The width varies in the range of 0.374 -- 0.462 {\AA}, with the standard deviations ranging from 0.014 to 0.030 {\AA}. It can be seen that the K$_{2}$ width increases from the equator to the poles, with a small peak at 0$\degree$. As the trend in K$_{1}$ width, the K$_{2}$ widths are smaller in the northern hemisphere, and the increase is steeper in the southern hemisphere, moving from 50$\degree$S to 70$\degree$S. The averaged K$_{2}$ width over all latitudes varies by 35.26\% between the minimum and maximum phases.

\subsection{Variations in K$_{3}$ Intensities} \label{subsec:k3int}
\begin{figure}
	\centering
	\setkeys{Gin}{draft=False}
	\includegraphics[width=\hsize]{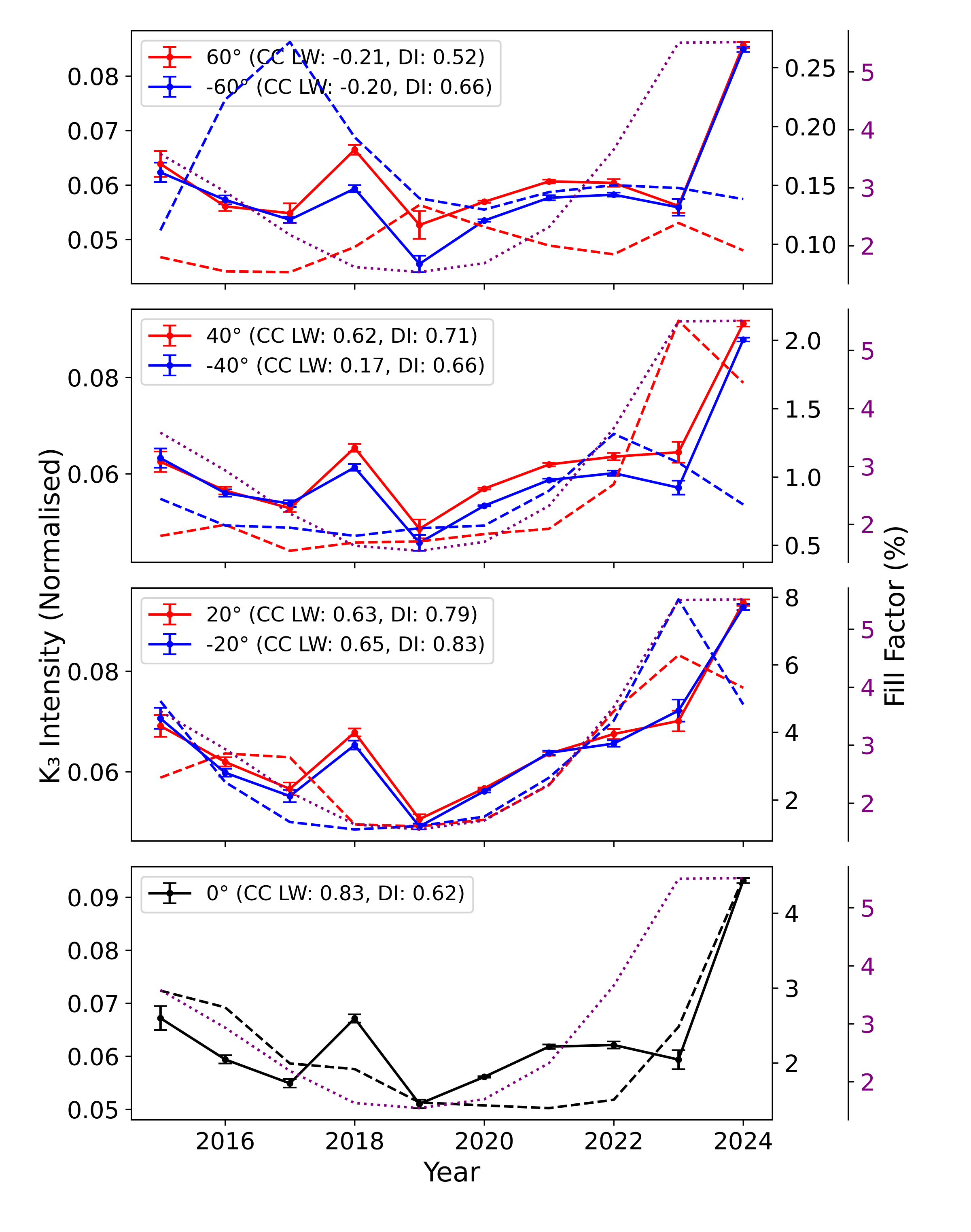}
	\caption{Temporal Variations in K$_{3}$ intensities for select latitudes. Circular points with error bars show the yearly-averaged intensities, red for northern latitudes and blue for southern. The red and blue (black for 0$\degree$) dashed lines show the yearly-averaged LW plage+spot FF for the respective hemispheres. The purple dotted line shows the yearly-averaged DI-FF.}
	\label{temp_var_k3}
\end{figure}

\begin{figure}
	\centering
	\setkeys{Gin}{draft=False}
	\includegraphics[width=\hsize]{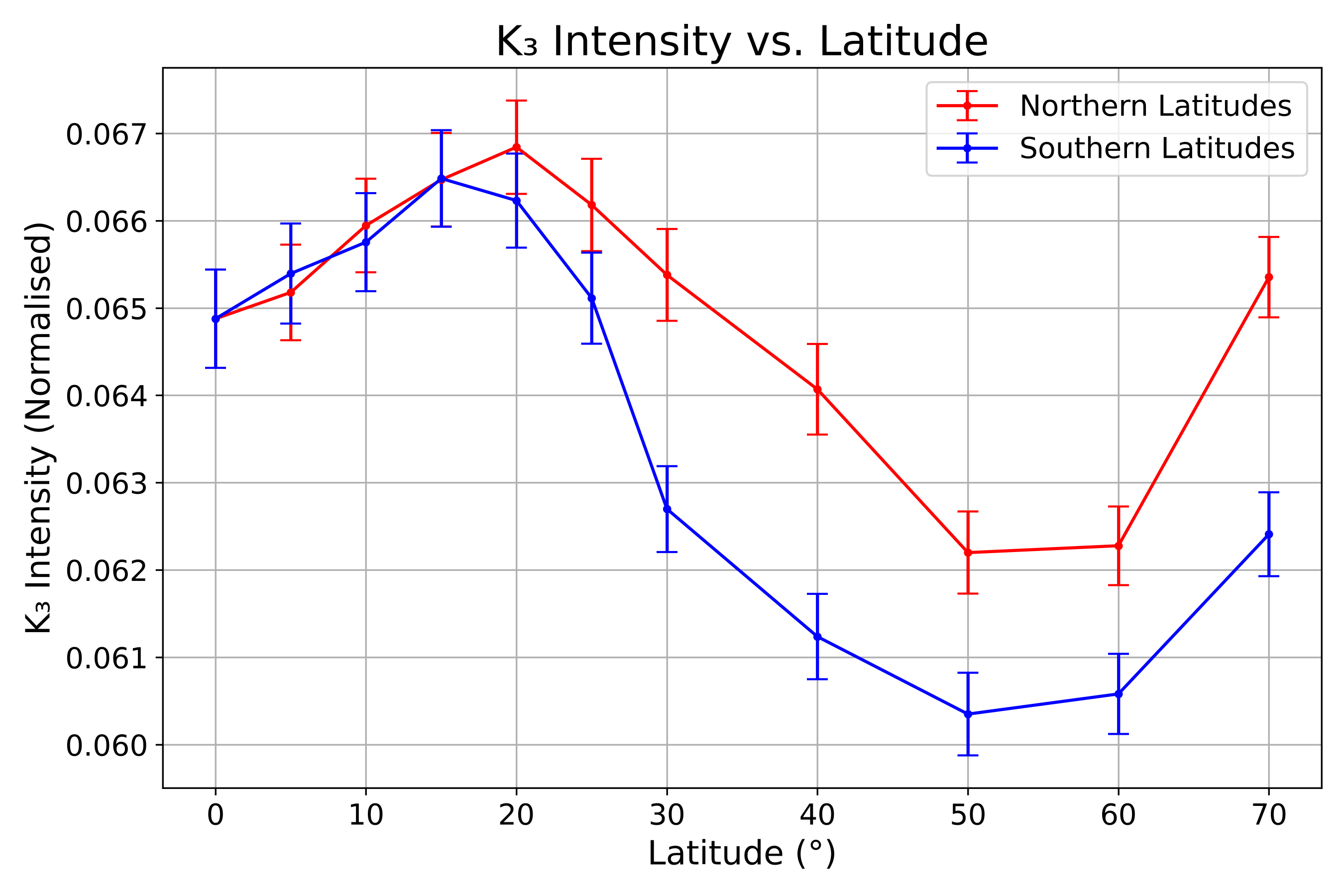}
	\caption{Time-averaged K$_{3}$ intensity variation as a function of latitude. Northern hemisphere in red and southern hemisphere in blue.}
	\label{k3var}
\end{figure}

Figure \ref{temp_var_k3} shows the variations in the K$_{3}$ intensities over the past 10 years for select latitudes. A similar trend is seen in each of the latitudes plotted, which is a decrease in the intensities during 2015 -- 2019/2020, followed by an increase towards 2024. The CCs between the K$_{3}$ intensity and LW-FF as well as DI-FF are mostly positive. The maximum LW- and DI-FF correlations of 0.83 and 0.83 are obtained at 0$\degree$ and 20$\degree$S latitudes, and minimum correlations of 0.17 and 0.52 are obtained at 40$\degree$S and 60$\degree$N latitudes. The negative CC of LW-FF at 60$\degree$ latitudes may be due to the minute noise in the data despite the outlier removal. A possibility for the anomalous behaviour of the parameter plots in 2018 could be the underperformance of CCD. \par
Figure \ref{k3var} shows the plot of K$_{3}$ intensity, averaged over time, as a function of latitude. The intensity varies in the range of 0.060 -- 0.067 units, with the standard deviations ranging from 0.012 to 0.014 units. The intensity variation shows an increase on either side of the 0$\degree$ latitude until 20$\degree$N and 15$\degree$S, thereupon the trend falls below the intensity value at 0$\degree$, moving towards 50$\degree$ on either side. From 50$\degree$ on either side (with the southern side lower than the northern side), the intensity then increases again towards the poles, higher in the northern hemisphere. The averaged K$_{3}$ intensity over all latitudes varies by 62.60\% between the minimum and maximum phases.

\subsection{Variations in K$_{2v}$/K$_{2r}$ Intensity Ratios} \label{subsec:k2vrratio}
\begin{figure}
	\centering
	\setkeys{Gin}{draft=False}
	\includegraphics[width=\hsize]{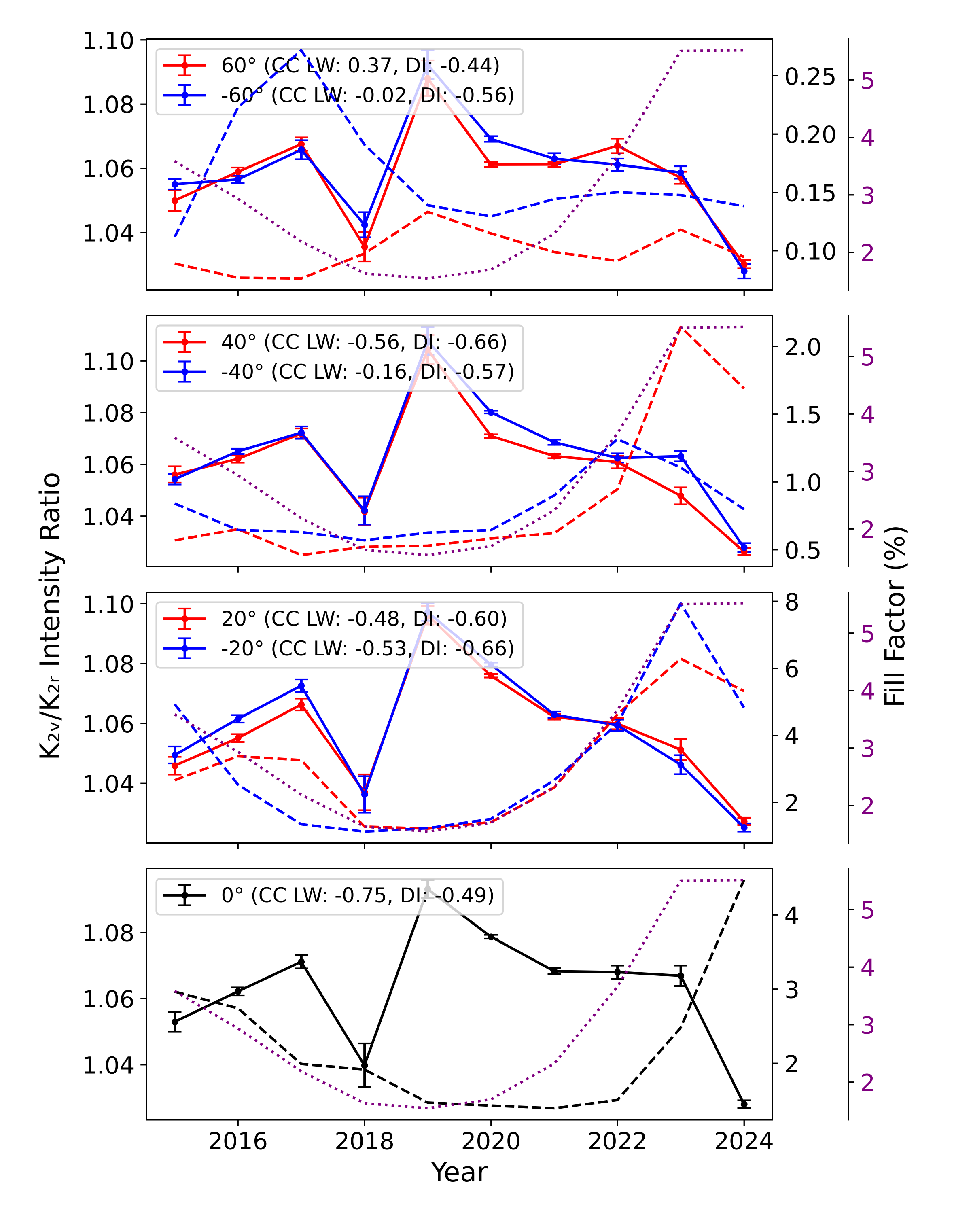}
	\caption{Temporal Variations in K$_{2v}$/K$_{2r}$ intensity ratios for select latitudes. Circular points with error bars show the yearly-averaged intensity ratios, red for northern latitudes and blue for southern. The red and blue (black for 0$\degree$) dashed lines show the yearly-averaged LW plage+spot FF for the respective hemispheres. The purple dotted line shows the yearly-averaged DI-FF.}
	\label{temp_var_k2vr}
\end{figure}

\begin{figure}
	\centering
	\setkeys{Gin}{draft=False}
	\includegraphics[width=\hsize]{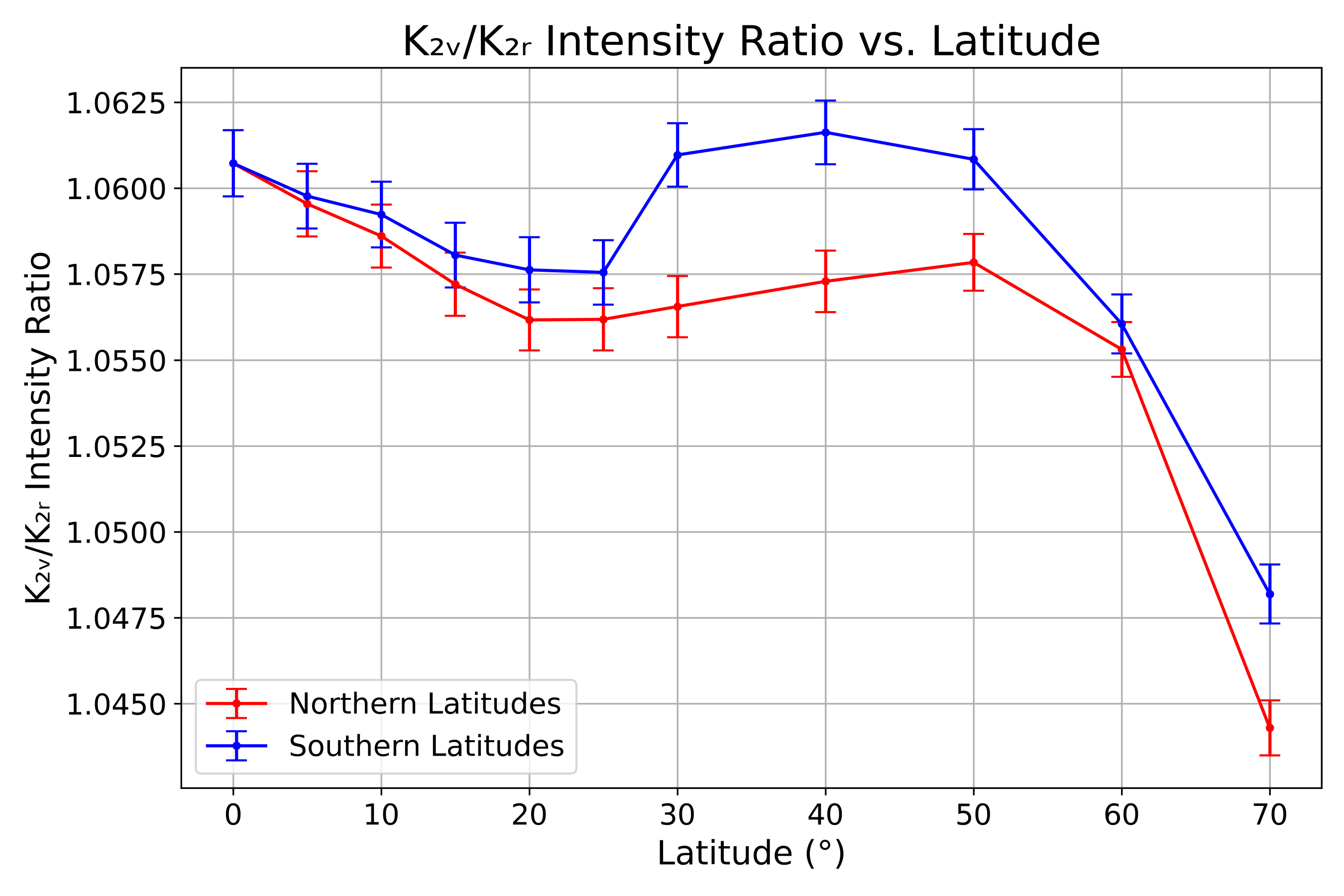}
	\caption{Time-averaged K$_{2v}$/K$_{2r}$ intensity ratio variation as a function of latitude. Northern hemisphere in red and southern hemisphere in blue.}
	\label{k2vrvar}
\end{figure}

Figure \ref{temp_var_k2vr} shows the variations in the K$_{2v}$/K$_{2r}$ intensity ratios over time for select latitudes. The intensity ratio for each of the latitudes plotted shows a similar trend, which is an increase in the ratios during 2015 -- 2019/2020, followed by a decrease towards 2024. The CCs between the K$_{2v}$/K$_{2r}$ intensity ratio and LW-FF as well as DI-FF are mostly negative. The maximum LW- and DI-FF negative correlations of -0.75 and -0.66 are obtained at 0$\degree$ and 20$\degree$S (also at 40$\degree$N for DI-FF), and minimum negative correlations of -0.02 and -0.44 are obtained at 60$\degree$S and 60$\degree$N latitudes. The anomalous behaviour of the parameter plot follows the same reasoning as the K$_{3}$ intensity. The small LW-FF correlation at 60$\degree$S and the positive LW-FF correlation at 60$\degree$N may be caused due to the anomalous behaviour. \par
Figure \ref{k2vrvar} shows the plot of the K$_{2v}$/K$_{2r}$ intensity ratio, averaged over time, as a function of latitude. The ratio varies in the range of 1.044 -- 1.062, with the standard deviations ranging from 0.017 to 0.020. From a central peak at 0$\degree$, the ratio falls on either side until 25$\degree$S and 20$\degree$N, whereupon the ratio rises (slightly above the ratio value at 0$\degree$, on the southern side), moving towards 40$\degree$S and 50$\degree$N. From these latitudes on either side, the ratio dips sharply towards the poles, steeper in the northern hemisphere. The ratios in the northern hemisphere are generally lower than those in the southern hemisphere. The averaged K$_{2v}$/K$_{2r}$ intensity ratio over all latitudes varies by 14.49\% between the minimum and maximum phases.

\subsection{Variations in K$_{2v}$/K$_{3}$ Intensity Ratios} \label{subsec:k2v3ratio}
\begin{figure}
	\centering
	\setkeys{Gin}{draft=False}
	\includegraphics[width=\hsize]{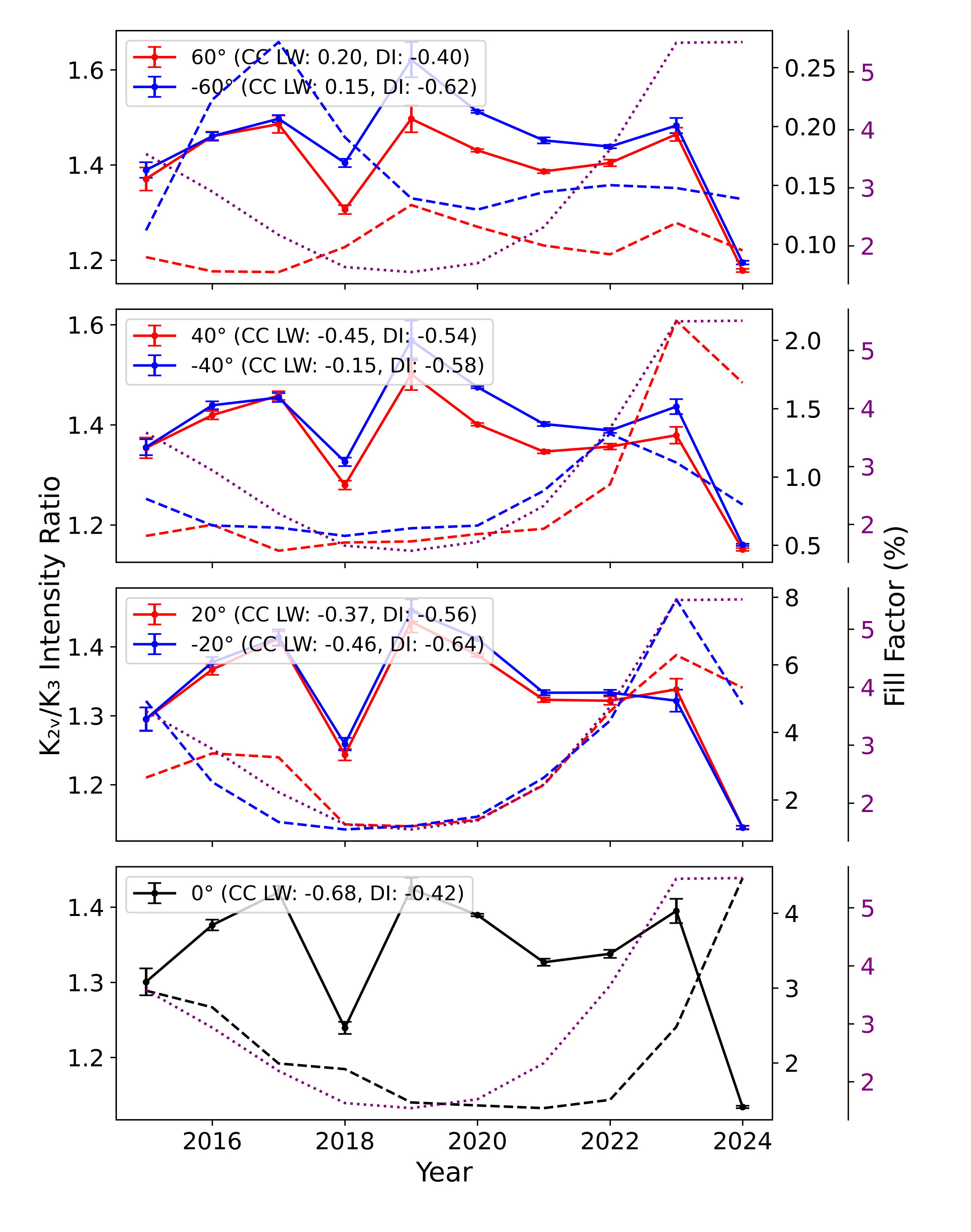}
	\caption{Temporal Variations in K$_{2v}$/K$_{3}$ intensity ratio for select latitudes. Circular points with error bars show the yearly-averaged intensity ratios, red for northern latitudes and blue for southern. The red and blue (black for 0$\degree$) dashed lines show the yearly-averaged LW plage+spot FF for the respective hemispheres. The purple dotted line shows the yearly-averaged DI-FF.}
	\label{temp_var_k2v3}
\end{figure}

\begin{figure}
	\centering
	\setkeys{Gin}{draft=False}
	\includegraphics[width=\hsize]{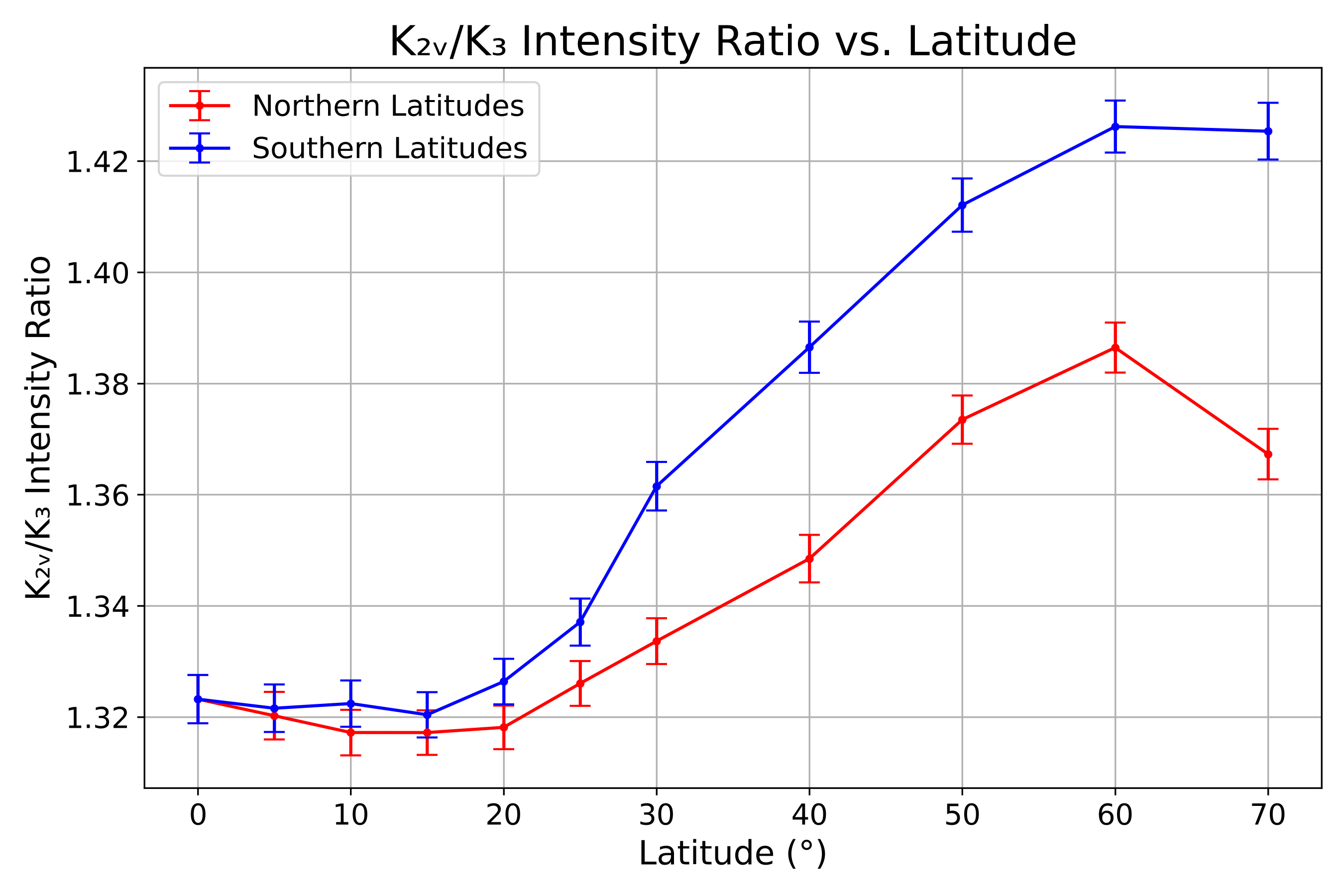}
	\caption{Time-averaged K$_{2v}$/K$_{3}$ intensity ratio variation as a function of latitude. Northern hemisphere in red and southern hemisphere in blue.}
	\label{k2v3var}
\end{figure}

Figure \ref{temp_var_k2v3} shows the variations in the K$_{2v}$/K$_{3}$ intensity ratios between 2015 and 2024 for select latitudes. The intensity ratios for all latitudes show similar trends, which are an increase during 2015 -- 2019/2020, followed by a decrease towards 2024. The CCs between the K$_{2v}$/K$_{3}$ intensity ratio and LW-FF as well as DI-FF are mostly negative. The maximum LW- and DI-FF negative correlations of -0.68 and -0.64 are obtained at 0$\degree$ and 20$\degree$S, and minimum negative correlations of -0.15 and -0.40 are obtained at 40$\degree$S and 60$\degree$N. The anomalous behaviour of the parameter plot follows the same reasoning as the K$_{3}$ intensity. The overall small LW-FF correlation and the positive LW-FF correlation at 60$\degree$ latitudes may be caused due to the anomalous behaviour. \par
Figure \ref{k2v3var} shows the plot of the K$_{2v}$/K$_{3}$ intensity ratio, averaged over time, as a function of latitude. The ratio varies in the range of 1.317 -- 1.426, with the standard deviations ranging from 0.100 to 0.126. From a small central bump at the 0$\degree$ latitude, the ratio rises on either side until 60$\degree$, thereupon, the ratio falls slightly towards the pole in the southern hemisphere, and the ratio falls sharply towards the pole in the northern hemisphere. The ratios in the northern hemisphere are generally lower than those in the southern hemisphere, and the rise in the southern hemisphere is steeper than that in the northern hemisphere. The averaged K$_{2v}$/K$_{3}$ intensity ratio over all latitudes varies by 36.52\% between the minimum and maximum phases.

\subsection{Variations in Flux Difference per Filling Factor Difference} \label{subsec:fluxdiff_ffdiff}
\begin{figure}
	\centering
	\setkeys{Gin}{draft=False}
	\includegraphics[width=\hsize]{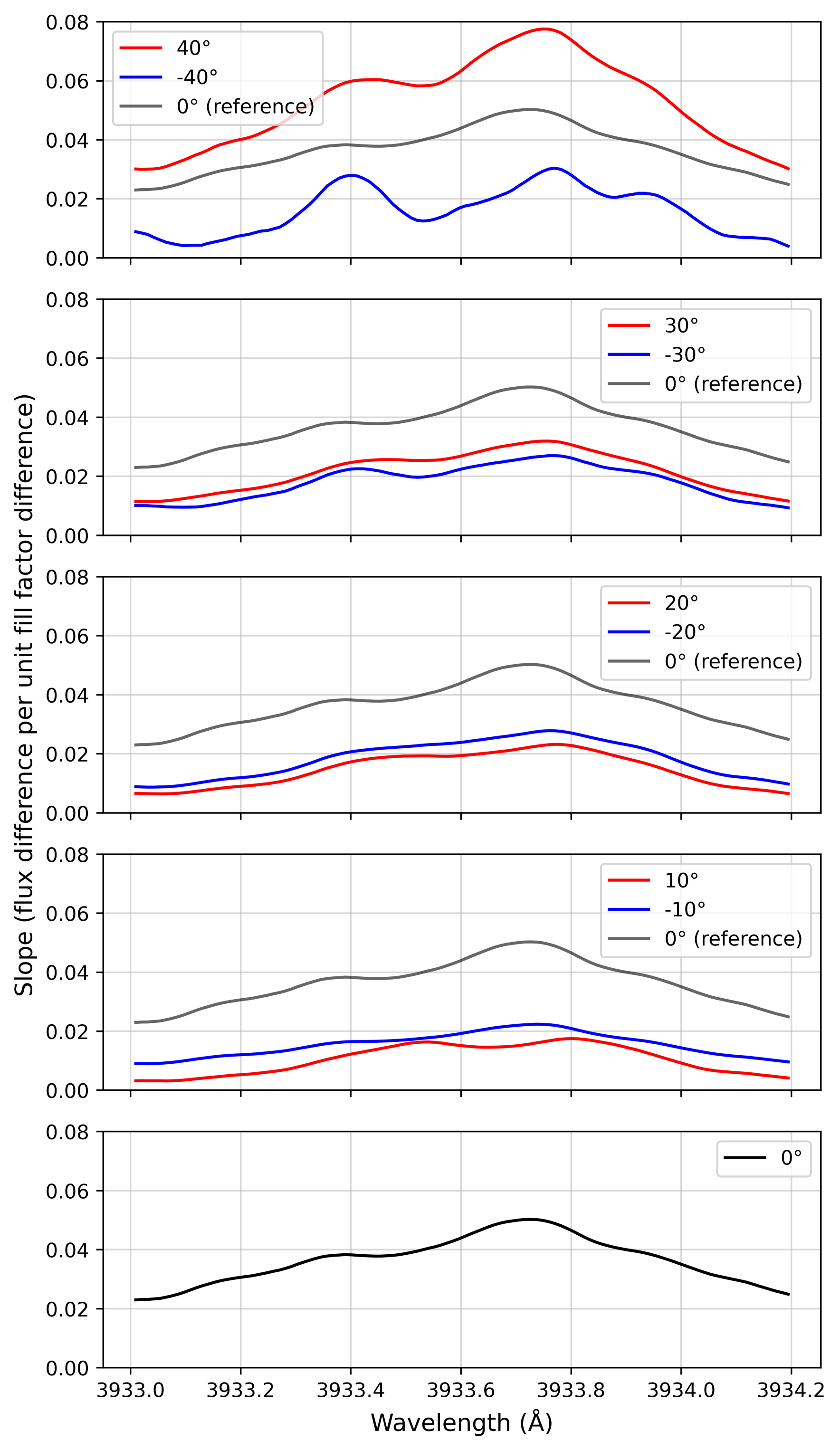}
	\caption{Flux difference per unit filling factor difference at various wavelengths. Northern hemisphere in red and southern hemisphere in blue. The 0$\degree$ latitude is shown as a reference with the other latitude pairs.}
	\label{flux_ff_diff}
\end{figure}

Figure \ref{flux_ff_diff} shows the flux difference per unit FF difference at wavelengths between 3933.0--3934.2 {\AA} for various latitudes. All profiles exhibit positive slopes, confirming that increased magnetic filling factor correlates with enhanced Ca II K line intensity. Each profile displays a prominent peak near the line core (3933.6--3933.7 {\AA}), indicating maximum sensitivity to activity changes occurs at the K$_{3}$ region. The ±40$\degree$ latitude pair shows the highest peak magnitudes, with the northern hemisphere reaching $\approx$0.075 and southern hemisphere $\approx$0.03, demonstrating clear north-south asymmetry. The ±30$\degree$ pair exhibits moderate asymmetry favouring the north with peaks of $\approx$0.032 (north) and $\approx$0.027 (south). The ±20$\degree$ pair displays reversed asymmetry with the southern hemisphere dominant at $\approx$0.028 compared to the northern hemisphere's $\approx$0.023. The ±10$\degree$ pair shows the southern hemisphere reaching $\approx$0.023 while the northern hemisphere peaks at $\approx$0.017. The equatorial belt (0$\degree$) shows an intermediate value of $\approx$0.05. The north-south asymmetries vary systematically with latitude, indicating latitude-dependent magnetic activity patterns during our observation period.

\section{Discussion} \label{sec:discussion}

The variations in K$_{1}$ and K$_{2}$ widths of the Ca-K line profile can provide insights into the magnetic field variations and activity of the sun \citep{2016SoPh..291.2967B}. The increase in temperature with height in the sun's chromosphere causes an increase in the widths of the Ca-K lines with latitude, which is seen as we move from the equator to the poles (\citealt{1960ApJ...132..202S}; \citealt{1966ApNr...10..101E}; \citealt{2023A&A...671A.130P}). This relationship highlights the non-uniformity of solar activity across latitudes, indicating that magnetic field strengths and temperatures vary significantly, which is important for understanding solar dynamics and predicting solar phenomena \citep{2015NatCo...6.6491M}. Another possibility could be due to increasing optical depth as we move towards the poles. Further, the K$_{1}$ width of the Ca-K line increases with solar activity, likely due to the higher temperature of active regions causing broadening of the Ca-K profile. Conversely, the K$_{2}$ width decreases with increasing solar activity. These findings emphasise the importance of monitoring these widths as indicators of solar activity levels and magnetic field variations. This has been previously reported by \citet{1981ApJ...249..798W}, \citet{1987ApJ...313..456S} and \citet{2015JApA...36...81S}. \citet{2024MNRAS.527.2940C} report their observations for the Meudon spectroheliogram datacubes, where they found the emission profiles to be thinner in the sunspot areas compared to plage areas, though they remarked that the result may not be universal. These trends are observed towards the 15-20$\degree$ latitudes on both hemispheres, as depicted by the width plots in Figures \ref{k1var} and \ref{k2var}, suggesting activities in those regions. These variations are not only significant for understanding solar dynamics but also hold the potential for improving space weather prediction models \citep{2021LRSP...18....4T}. \par

Increased K$_{3}$ intensity suggests a higher abundance of calcium ions in the solar atmosphere \citep{2000SoPh..195..269G}. Changes in the K$_{3}$ intensity from the equator to the poles can unveil latitudinal dependencies, as these variations in intensity could reflect the variations in solar activity across latitudes. Comparing with K$_{1}$ and K$_{2}$ widths, we see a correlated trend with K$_{3}$ intensity in Figure \ref{k3var}, where the intensity increases towards the 15-20$\degree$ latitudes on both hemispheres, again suggesting activities in those areas. Whereas, the decrease in the intensity between 20$\degree$ and 50$\degree$ on either hemisphere may point to fewer activities occurring in these latitudes. This asymmetry between hemispheres may provide insights into underlying mechanisms driving hemispheric solar activity \citep{2014SSRv..186..251N}. Further, the increase in intensity beyond these minima may be due to higher temperatures in the polar regions as compared with equatorial regions (\citealt{1955VA......1..658D}; \citealt{1966MNRAS.131..407P}; \citealt{1960BAN....15...85B}). The higher intensity in the northern hemisphere suggests a higher amount of activity. This suggests that monitoring calcium ion abundance can serve as an effective proxy for assessing regional solar activity. \par

The K$_{2v}$/K$_{2r}$ intensity ratio provides the comparison of the violet (K$_{2v}$) and red (K$_{2r}$) components of the K$_{2}$ peaks in the Calcium-K spectrum. Changes in this ratio primarily reflect velocity fields in the chromosphere, with asymmetries largely resulting from Doppler shifts (\citealt{1974SoPh...37...85G}; \citealt{1983ApJ...272..355C}). The presence of upward-propagating acoustic shocks in the chromosphere tends to enhance emission in the violet wing (K$_{2v}$) relative to the red wing (K$_{2r}$), producing profiles with K$_{2v}$/K$_{2r}$ > 1 in most observations \citep{1997ApJ...481..500C}. This asymmetry emerges because shocks create large velocity gradients that shift the absorption profiles of material behind the shock towards shorter wavelengths, reducing opacity on the violet side of the line while increasing it on the red side \citep{1997ApJ...481..500C}. \citet{1971SoPh...17..316B} suggest that changes in the K$_{3}$ intensity can cause variations in the K$_{2v}$ and K$_{2r}$ intensities and that higher absorption (the K$_{3}$ core) may result in a decrease in K$_{2v}$ intensity compared to K$_{2r}$ intensity. Understanding this ratio may be important for interpreting chromospheric dynamics and linking them to broader magnetic activity patterns. We observe that the peaks and dips of the K$_{2v}$/K$_{2r}$ intensity ratio in Figure \ref{k2vrvar} correspond to the dips and peaks of K$_{3}$ intensity in Figure \ref{k3var}. These figures confirm the above-mentioned suggestion by \citet{1971SoPh...17..316B}. \par

The K$_{2v}$/K$_{3}$ intensity ratio reveals information about the relative strength of the emissions from the lower chromosphere and the upper chromosphere (\citealt{1998ASPC..140..293W}, \citealt{1988KodOB...9..159S}). An increased K$_{2v}$/K$_{3}$ ratio might suggest enhanced temperatures or specific conditions leading to stronger violet emission relative to the overall intensity at the K$_{3}$ central minimum. This finding emphasises how temperature gradients influence chromospheric emissions, contributing to our understanding of solar atmospheric structure. These temperature gradients may also reflect changes in the magnetic field structure, which could be important for future magnetic reconstructions of the solar atmosphere \citep{2017SSRv..210...37L}. From Figure \ref{k2v3var}, we can observe that the overall trend of the K$_{2v}$/K$_{3}$ intensity ratio increases from the equator to the poles. As suggested earlier, the temperatures in the polar regions are higher compared to equatorial regions. The K$_{2v}$ line is stronger at higher temperatures, whereas the K$_{3}$ line represents the absorption core in cooler regions. \par

The K$_{1}$ and K$_{2}$ widths of the Ca-K line profiles have temporal variations that are dependent on the solar cycle. The K$_{1}$ width increases/decreases towards the solar maximum/minimum, whereas the K$_{2}$ width decreases/increases. These trends are apparent from Figures \ref{temp_var_k1} and \ref{temp_var_k2}. The 20$\degree$ latitudes on both hemispheres display larger K$_{1}$ width variations in comparison to the other latitudes as more activities occur around the 20$\degree$ belt. This is due to the equatorial drift of the sunspots from around 40$\degree$ to the equator. The sunspots being magnetic activities contribute to the increased K$_{1}$ widths. The variations at 40$\degree$ and the equator are lower than the latitudes in between. This is because the number of sunspots is lower when the cycle starts around 40$\degree$ latitudes. It then increases as the cycle progresses with the sunspots drifting towards the equator and finally reduces as the spots approach the equator. The strong positive correlations between K$_{1}$ width and both plage and spot filling factors (with maximum correlations of 0.95 and 0.96 at 20$\degree$N and 40$\degree$N, respectively) confirm that this parameter is directly influenced by magnetic activity. The notably weaker correlations at high latitudes (minimum of 0.35 and 0.33 at 60$\degree$S) align with expectations from the Butterfly diagram pattern, where active regions are concentrated between ±40$\degree$ latitude and migrate equatorwards as the cycle progresses. As with the K$_{1}$ width, the K$_{2}$ width displays larger variations in the 20$\degree$ latitudes than those in the other latitudes. Interestingly, K$_{2}$ width shows strong negative correlations with both plage and spot filling factors (maximum negative correlations of -0.93 and -0.90 at 20$\degree$S), indicating an inverse relationship with surface magnetic activity. As the solar cycle progresses, the activities increase, leading to an increase in the intensities with the magnetic field strengths. This may have contributed to the smaller K$_{2}$ widths towards the solar maximum and larger widths towards the solar minimum. The weaker correlation at the equator (minimum negative correlations of -0.21 and -0.05) suggests that this parameter may be influenced by additional factors beyond simple active region presence. Further, this indicates that specific latitudinal bands are more sensitive to changes in solar activity, which is critical for modelling solar cycles and predicting related phenomena. Continued study of these latitudinal variations could help refine our understanding of the relationship between magnetic activity and the solar cycle \citep{2015LRSP...12....4H}. \par

The K$_{3}$ intensities of the Ca-K line profiles also have temporal variations that follow the solar cycles (Figure \ref{temp_var_k3}). The K$_{3}$ intensity increases/decreases towards the solar maximum/minimum. This parameter shows consistent positive correlations with both plage and spot filling factors across all latitudes (maximum correlations of 0.81 and 0.73 at 20$\degree$S), confirming that K$_{3}$ intensity directly responds to the presence of magnetic activity. As the activities increase, the energy released increases, and thus, the observed intensity increases. In contrast, as seen in the latitude-wise variations, the K$_{2v}$/K$_{2r}$ (Figure \ref{temp_var_k2vr}) and K$_{2v}$/K$_{3}$ (Figure \ref{temp_var_k2v3}) intensity ratios show the opposite behaviour to the K$_{3}$ intensity. That is, the ratios increase/decrease towards the solar minimum/maximum. The anomalous behaviour observed in 2018, particularly in the intensity and their ratio parameters, is not clearly understood. One possibility could be the underperformance of CCD. These observations highlight the intricate relationships between different emissions and their dependence on solar cycle phases, emphasising their importance for in-depth modelling efforts. \par

When analysing our data in the context of the Butterfly diagram, we observe that the strongest correlations between line profile parameters and magnetic activity indicators (plage and spot filling factors) consistently appear within ±40$\degree$ latitude, which precisely corresponds to the active region belt where the Butterfly pattern manifests. The temporal evolution of these correlations further supports this connection, with signals at higher latitudes (e.g., 40$\degree$N) being more pronounced during the ascending phase of the cycle, consistent with the polewards migration of active regions. The weakening correlations beyond ±40$\degree$ latitude is due to the lack of spots and plages in these regions. \par

As expected, the plage and spot CCs for each parameter confirm these temporal behaviours. In each parameter, the solar cycle dependence appears to be stronger between 20$\degree$N and 20$\degree$S, slightly weaker towards 40$\degree$N and 40$\degree$S, and weakest towards the poles (with CCs of K$_{2}$ width and K$_{2v}$/K$_{3}$ intensity ratio at 60$\degree$S as exceptions). Generally, the correlations seem to be weaker in the northern hemisphere than in the southern hemisphere, except for the K$_{1}$ width and the 40$\degree$ belts of K$_{2}$ width, the K$_{3}$ intensity and the K$_{2v}$/K$_{2r}$ intensity ratio, where northern hemisphere correlations are stronger. This hemispheric asymmetry in correlation strength may indicate fundamental differences in magnetic field evolution between the northern and southern hemispheres during this phase of the solar cycle. For each parameter, except for the K$_{1}$ width, the 20$\degree$S latitude shows the highest correlations compared to other latitudes. The plage and spot CCs for K$_{1}$ width at 20$\degree$S are 0.92 each, whereas those of 20$\degree$N are 0.95 each, despite which our observation that the 20$\degree$S latitudes show the highest correlation holds true. Further, in terms of absolute CCs, K$_{1}$ width shows higher correlations than other parameters. The particularly strong correlations between K$_{1}$ width and magnetic activity indicators suggest that this parameter may serve as the most reliable chromospheric proxy for tracking solar magnetic activity evolution. This information may be important for developing predictive models regarding solar behaviour and its implications for space weather. \par

The segmentation of SDO data to extract sunspot and plage filling factors has proven valuable in quantifying the relationship between our spectral parameters and surface magnetic features. The strong latitude-dependent correlations confirm that variations in Ca-K line profiles are directly linked to the changing distribution of active regions over the solar cycle, rather than being merely coincidental with the sunspot number trend. The filling factor analysis particularly reveals how the type and coverage of magnetic features at different latitudinal bands influence specific spectral parameters in distinct ways, providing a more mechanistic understanding of the observed variations beyond simple cycle-dependent correlations. \par

Previous work by \citet{1987ApJ...313..456S} showed the temporal variations in K$_{3}$ intensity and the intensity ratios of K$_{2v}$/K$_{3}$ and K$_{2v}$/K$_{2r}$ between the years 1968 and 1984. \citet{2015IIA...PhD...Thesis} obtained the temporal variations in K$_{1}$ and K$_{2}$ widths as well as the latitudinal variations (during quiet and active phases) for all five parameters between the years 1989 to 2011. The results from our study conform to the trends seen historically in their work. This consistency supports the reliability of these trends over multiple solar cycles, suggesting that these parametric variations are persistent features of solar activity. Such agreement highlights the importance of continued observations to deepen our understanding of these long-term solar behaviours. Future comparisons with additional datasets could help validate these findings over even longer periods, strengthening our models of solar behaviour \citep{2020A&A...639A..88C}. \par

The slope profiles in Figure \ref{flux_ff_diff} represent the mean spectral response of the chromosphere to magnetic activity, effectively showing how the Ca II K line shape changes when active regions are added to a quiet Sun background. The consistent peak near the K$_{3}$ line core reflects the well-established behaviour where active regions fill in and brighten the central reversal, attributed to enhanced temperatures in the upper chromosphere of magnetic features \citep{1974SoPh...39...49S}. The latitude dependence of these profiles likely reflects the combination of several factors, such as the intrinsic distribution of active regions over the solar cycle, centre-to-limb viewing effects that systematically affect different latitude bands, and potential variations in the magnetic field properties at different latitudes. The methodology follows similar approaches described by \citet{2024MNRAS.527.2940C}, who noted that such latitude-dependent intensity profiles inherently include centre-to-limb variation effects on active region emission. The hemispheric differences observed in our data period (2015-2024), which spans the transition from Solar Cycle 24 to 25, reflect the latitude-dependent activity patterns characteristic of this phase of the solar cycle. The spectrally resolved approach provides more detailed information than traditional line indices, complementing our findings that K$_{1}$ width correlates with filling factor by the precise mapping of wavelengths which drive these correlations.

\section{Conclusion} \label{sec:conclusion}

We obtained observations of the Ca-K line profiles of the sun for a period of about ten years from 2015 to 2024 at the Kodaikanal Solar Observatory. From these line profiles, we have extracted the K$_{1}$ width, K$_{2}$ width, K$_{3}$ intensity and the intensity ratios of K$_{2v}$/K$_{2r}$ and K$_{2v}$/K$_{3}$ as functions of latitude and time. As we move towards the polar regions, the time-averaged K$_{1}$ and K$_{2}$ widths and intensity ratio of K$_{2v}$/K$_{3}$ increase, whereas the time-averaged K$_{3}$ intensity and intensity ratio of K$_{2v}$/K$_{2r}$ decrease. The parameters, K$_{1}$ width and K$_{3}$ intensity, show peaks around 15-20$\degree$ on both hemispheres, whereas K$_{2}$ width and the intensity ratios of K$_{2v}$/K$_{2r}$ and K$_{2v}$/K$_{3}$ show dips around those latitudes. Generally, the trends indicate that the activity regions are mostly found between 40$\degree$S and 40$\degree$N. Finally, the yearly-averaged temporal variation plots show that the K$_{1}$ width and K$_{3}$ intensity increase/decrease with solar maximum/minimum, whereas the K$_{2}$ width and the intensity ratios of K$_{2v}$/K$_{2r}$ and K$_{2v}$/K$_{3}$ show decrease/increase. These temporal behaviours are further confirmed by the correlation coefficients between the parameters and the plage and spot filling factors. Our results are also in agreement with the previous spatial and temporal studies between 1986 and 2011. Additionally, we derived slope profiles representing the spectral response of the Ca II K line to changes in magnetic filling factor across different latitudes. These profiles consistently peak near the K$_{3}$ line core, confirming that the primary spectral signature of solar activity occurs through filling-in of the central reversal. The profiles also show north-south differences that vary with latitude. Further studies are on the anvil to understand the full significance of the results.

\section*{Acknowledgements} \label{sec:acknowledgments}
The authors declare no conflict of interest. The authors gratefully acknowledge the tireless effort of the KSO team to obtain the spectral data and the IIA scientists for developing the calibration techniques for the data. This research has made use of data from the HMI instrument on board the SDO, a mission for NASA's Living With a Star program. The authors acknowledge the use of the \textsc{SolAster}\footnote{\url{https://github.com/tamarervin/SolAster}} Python package for data analysis. Some of the authors also express their gratitude to The Chancellor of Amrita Vishwa Vidyapeetham for their support. This work is funded by the Department of Science and Technology, Government of India.

\section*{Data Availability}

The raw Calcium II K spectral data are available on request at the Indian Institute of Astrophysics. HMI data used in this study are publicly available from the \textsc{Joint Science Operations Center}\footnote{\url{http://jsoc.stanford.edu/}} (JSOC) at Stanford University. The Mg II index data are available from the \textsc{UVSAT Group}\footnote{\url{https://www.iup.uni-bremen.de/UVSAT/}} of the University of Bremen.



\bibliographystyle{mnras}

\begin{thebibliography}{}
	\bibitem[\protect\citeauthoryear{Bappu \& Sivaraman}{1971}]{1971SoPh...17..316B} Bappu M.~K.~V., Sivaraman K.~R., 1971, SoPh, 17, 316. doi:10.1007/BF00150035
	\bibitem[\protect\citeauthoryear{Beckers}{1960}]{1960BAN....15...85B} Beckers J.~M., 1960, BAN, 15, 85
	\bibitem[\protect\citeauthoryear{Bertello et al.}{2016}]{2016SoPh..291.2967B} Bertello L., Pevtsov A., Tlatov A., Singh J., 2016, SoPh, 291, 2967. doi:10.1007/s11207-016-0927-9
	\bibitem[\protect\citeauthoryear{Bertello, Pevtsov, \& Ulrich}{2020}]{2020ApJ...897..181B} Bertello L., Pevtsov A.~A., Ulrich R.~K., 2020, ApJ, 897, 181. doi:10.3847/1538-4357/ab9746
	\bibitem[\protect\citeauthoryear{Carlsson \& Stein}{1997}]{1997ApJ...481..500C} Carlsson M., Stein R.~F., 1997, ApJ, 481, 500. doi:10.1086/304043
	\bibitem[\protect\citeauthoryear{Charbonneau}{2020}]{2020LRSP...17....4C} Charbonneau P., 2020, LRSP, 17, 4. doi:10.1007/s41116-020-00025-6
	\bibitem[\protect\citeauthoryear{Chatzistergos et al.}{2020}]{2020A&A...639A..88C} Chatzistergos T., Ermolli I., Krivova N.~A., Solanki S.~K., Banerjee D., Barata T., Belik M., et al., 2020, A\&A, 639, A88. doi:10.1051/0004-6361/202037746
	\bibitem[\protect\citeauthoryear{Chatzistergos, Krivova, \& Ermolli}{2022}]{2022FrASS...938949C} Chatzistergos T., Krivova N.~A., Ermolli I., 2022, FrASS, 9, 336. doi:10.3389/fspas.2022.1038949
	\bibitem[\protect\citeauthoryear{Cram \& Dame}{1983}]{1983ApJ...272..355C} Cram L.~E., Dame L., 1983, ApJ, 272, 355. doi:10.1086/161299
	\bibitem[\protect\citeauthoryear{Cretignier, Pietrow, \& Aigrain}{2024}]{2024MNRAS.527.2940C} Cretignier M., Pietrow A.~G.~M., Aigrain S., 2024, MNRAS, 527, 2940. doi:10.1093/mnras/stad3292
	\bibitem[\protect\citeauthoryear{Das \& Abhyankar}{1955}]{1955VA......1..658D} Das A.~K., Abhyankar K.~D., 1955, VA, 1, 658. doi:10.1016/0083-6656(55)90080-5
	\bibitem[\protect\citeauthoryear{Engvold}{1966}]{1966ApNr...10..101E} Engvold O., 1966, ApNr, 10, 101
	\bibitem[\protect\citeauthoryear{Ervin et al.}{2022}]{2022ascl.soft07009E} Ervin T., Halverson S., Burrows A., Murphy N., Roy A., Haywood R.~D., Rescigno F., et al., 2022, ascl.soft. ascl:2207.009
	\bibitem[\protect\citeauthoryear{Froehlich et al.}{1991}]{1991suti.conf...11F} Froehlich C., Foukal P.~V., Hickey J.~R., Hudson H.~S., Willson R.~C., 1991, suti.conf, 11
	\bibitem[\protect\citeauthoryear{Fr{\"o}hlich \& Lean}{1998}]{1998GeoRL..25.4377F} Fr{\"o}hlich C., Lean J., 1998, GeoRL, 25, 4377. doi:10.1029/1998GL900157
	\bibitem[\protect\citeauthoryear{Gray et al.}{2010}]{2010RvGeo..48.4001G} Gray L.~J., Beer J., Geller M., Haigh J.~D., Lockwood M., Matthes K., Cubasch U., et al., 2010, RvGeo, 48, RG4001. doi:10.1029/2009RG000282
	\bibitem[\protect\citeauthoryear{Grigoryeva, Ozhogina, \& Teplitskaya}{2000}]{2000SoPh..195..269G} Grigoryeva S.~A., Ozhogina O.~A., Teplitskaya R.~B., 2000, SoPh, 195, 269. doi:10.1023/A:1005299727536
	\bibitem[\protect\citeauthoryear{Grossmann-Doerth, Kneer, \& von Uexk{\"u}ll}{1974}]{1974SoPh...37...85G} Grossmann-Doerth U., Kneer F., von Uexk{\"u}ll M., 1974, SoPh, 37, 85. doi:10.1007/BF00157846
	\bibitem[\protect\citeauthoryear{Hale \& Ellerman}{1904}]{1904ApJ....19...41H} Hale G.~E., Ellerman F., 1904, ApJ, 19, 41. doi:10.1086/141083
	\bibitem[\protect\citeauthoryear{Hathaway}{2015}]{2015LRSP...12....4H} Hathaway D.~H., 2015, LRSP, 12, 4. doi:10.1007/lrsp-2015-4
	\bibitem[\protect\citeauthoryear{Keil \& Worden}{1984}]{1984ApJ...276..766K} Keil S.~L., Worden S.~P., 1984, ApJ, 276, 766. doi:10.1086/161663
	\bibitem[\protect\citeauthoryear{Lagg et al.}{2017}]{2017SSRv..210...37L} Lagg A., Lites B., Harvey J., Gosain S., Centeno R., 2017, SSRv, 210, 37. doi:10.1007/s11214-015-0219-y
	\bibitem[\protect\citeauthoryear{Makarov \& Tlatov}{2000}]{2000ARep...44..759M} Makarov V.~I., Tlatov A.~G., 2000, ARep, 44, 759. doi:10.1134/1.1320502
	\bibitem[\protect\citeauthoryear{McIntosh et al.}{2015}]{2015NatCo...6.6491M} McIntosh S.~W., Leamon R.~J., Krista L.~D., Title A.~M., Hudson H.~S., Riley P., Harder J.~W., et al., 2015, NatCo, 6, 6491. doi:10.1038/ncomms7491
	\bibitem[\protect\citeauthoryear{Morrill, Floyd, \& McMullin}{2011}]{2011SoPh..269..253M} Morrill J.~S., Floyd L., McMullin D., 2011, SoPh, 269, 253. doi:10.1007/s11207-011-9708-7
	\bibitem[\protect\citeauthoryear{Nagaraju}{2008}]{2008IIA...PhD...Thesis} Nagaraju, K., 2008, PhD Thesis, Indian Institute of Astrophysics
	\bibitem[\protect\citeauthoryear{Nindos \& Zirin}{1998}]{1998SoPh..179..253N} Nindos A., Zirin H., 1998, SoPh, 179, 253. doi:10.1023/A:1005046114362
	\bibitem[\protect\citeauthoryear{Norton, Charbonneau, \& Passos}{2014}]{2014SSRv..186..251N} Norton A.~A., Charbonneau P., Passos D., 2014, SSRv, 186, 251. doi:10.1007/s11214-014-0100-4
	\bibitem[\protect\citeauthoryear{Oranje}{1983}]{1983A&A...124...43O} Oranje B.~J., 1983, A\&A, 124, 43
	\bibitem[\protect\citeauthoryear{Ortiz \& Rast}{2005}]{2005MmSAI..76.1018O} Ortiz A., Rast M., 2005, MmSAI, 76, 1018
	\bibitem[\protect\citeauthoryear{Pap et al.}{1997}]{1997ESASP.415..251P} Pap J.~M., Floyd L., Lee R.~B., Parker D., Puga L., Ulrich R., Varadi F., et al., 1997, ESASP, 415, 251
	\bibitem[\protect\citeauthoryear{Pietrow et al.}{2023}]{2023A&A...671A.130P} Pietrow A.~G.~M., Kiselman D., Andriienko O., Petit dit de la Roche D.~J.~M., D{\'\i}az Baso C.~J., Calvo F., 2023, A\&A, 671, A130. doi:10.1051/0004-6361/202244811
	\bibitem[\protect\citeauthoryear{Plaskett}{1966}]{1966MNRAS.131..407P} Plaskett H.~H., 1966, MNRAS, 131, 407. doi:10.1093/mnras/131.3.407
	\bibitem[\protect\citeauthoryear{Priyal et al.}{2023}]{2023ApJ...944..218P} Priyal M., Singh J., Ravindra B., Sindhuja G., 2023, ApJ, 944, 218. doi:10.3847/1538-4357/acaf60
	\bibitem[\protect\citeauthoryear{Sindhuja}{2015}]{2015IIA...PhD...Thesis} Sindhuja, G., 2015, PhD Thesis, Indian Institute of Astrophysics
	\bibitem[\protect\citeauthoryear{Sindhuja \& Singh}{2015}]{2015JApA...36...81S} Sindhuja G., Singh J., 2015, JApA, 36, 81. doi:10.1007/s12036-015-9330-4
	\bibitem[\protect\citeauthoryear{Sindhuja, Singh, \& Ravindra}{2014}]{2014ApJ...792...22S} Sindhuja G., Singh J., Ravindra B., 2014, ApJ, 792, 22. doi:10.1088/0004-637X/792/1/22
	\bibitem[\protect\citeauthoryear{Singh}{1988}]{1988KodOB...9..159S} Singh J., 1988, KodOB, 9, 159
	\bibitem[\protect\citeauthoryear{Sivaraman et al.}{1987}]{1987ApJ...313..456S} Sivaraman K.~R., Singh J., Bagare S.~P., Gupta S.~S., 1987, ApJ, 313, 456. doi:10.1086/164985
	\bibitem[\protect\citeauthoryear{Shine \& Linsky}{1974}]{1974SoPh...39...49S} Shine R.~A., Linsky J.~L., 1974, SoPh, 39, 49. doi:10.1007/BF00154970
	\bibitem[\protect\citeauthoryear{Solanki, Krivova, \& Haigh}{2013}]{2013ARA&A..51..311S} Solanki S.~K., Krivova N.~A., Haigh J.~D., 2013, ARA\&A, 51, 311. doi:10.1146/annurev-astro-082812-141007
	\bibitem[\protect\citeauthoryear{Smith}{1960}]{1960ApJ...132..202S} Smith E.~V.~P., 1960, ApJ, 132, 202. doi:10.1086/146913
	\bibitem[\protect\citeauthoryear{Temmer}{2021}]{2021LRSP...18....4T} Temmer M., 2021, LRSP, 18, 4. doi:10.1007/s41116-021-00030-3
	\bibitem[\protect\citeauthoryear{Vernazza, Avrett, \& Loeser}{1981}]{1981ApJS...45..635V} Vernazza J.~E., Avrett E.~H., Loeser R., 1981, ApJS, 45, 635. doi:10.1086/190731
	\bibitem[\protect\citeauthoryear{White et al.}{1998}]{1998ASPC..140..293W} White O.~R., Livingston W.~C., Keil S.~L., Henry T.~W., 1998, ASPC, 140, 293
	\bibitem[\protect\citeauthoryear{White \& Livingston}{1981}]{1981ApJ...249..798W} White O.~R., Livingston W.~C., 1981, ApJ, 249, 798. doi:10.1086/159338
	\bibitem[\protect\citeauthoryear{White \& Livingston}{1978}]{1978ApJ...226..679W} White O.~R., Livingston W., 1978, ApJ, 226, 679. doi:10.1086/156650
	\bibitem[\protect\citeauthoryear{White \& Suemoto}{1968}]{1968SoPh....3..523W} White O.~R., Suemoto Z., 1968, SoPh, 3, 523. doi:10.1007/BF00151934
	\bibitem[\protect\citeauthoryear{Worden \& Harvey}{2000}]{2000SoPh..195..247W} Worden J., Harvey J., 2000, SoPh, 195, 247. doi:10.1023/A:1005272502885


\end{thebibliography}




\appendix
\section{CDF Distributions} \label{sec:app_a}
\begin{figure}
	\centering
	\setkeys{Gin}{draft=False}
	\includegraphics[width=\hsize]{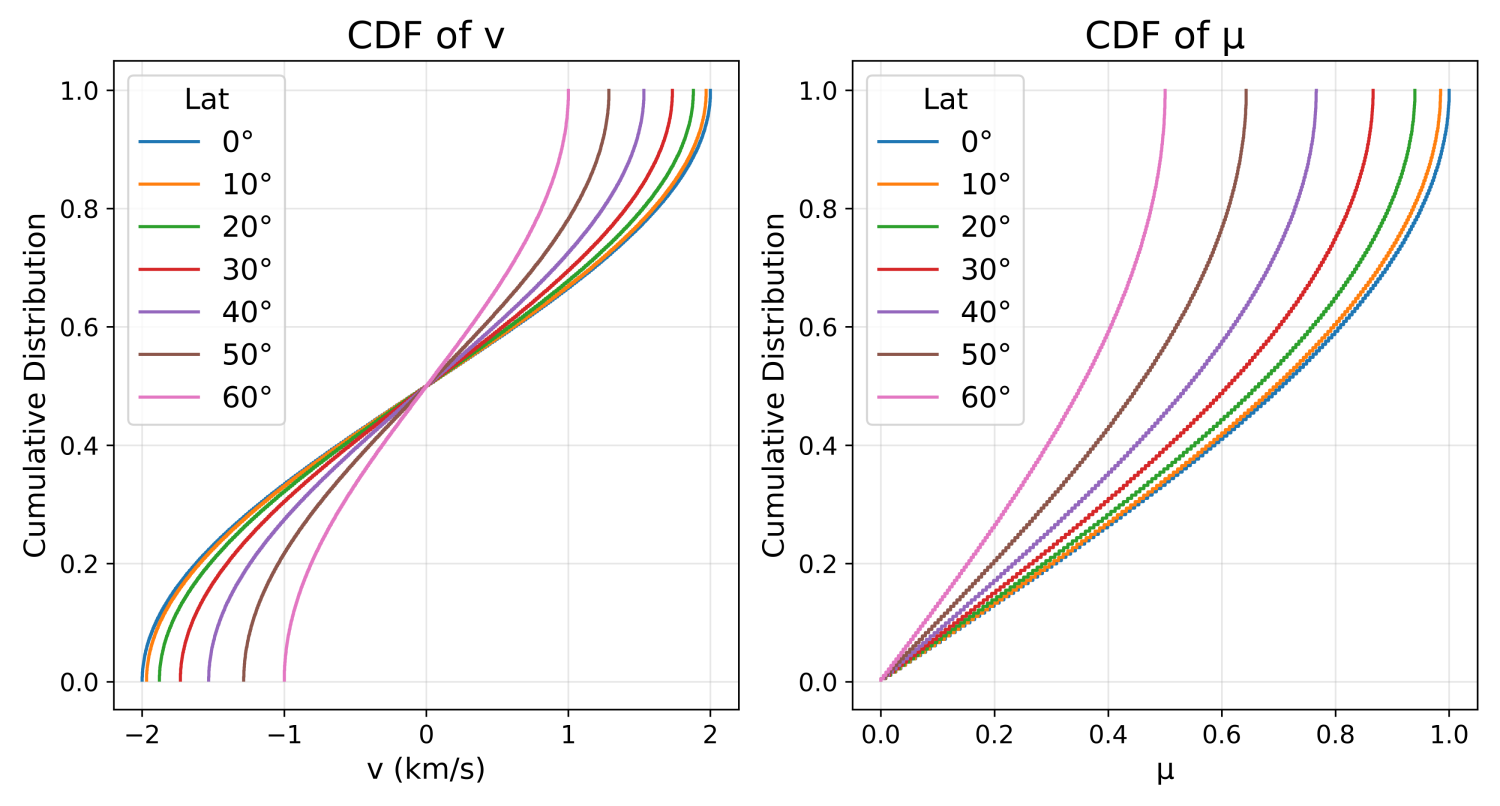}
	\caption{Cumulative distribution plots of (Left) $v$ and (Right) $\mu$ for various latitudes.}
	\label{cdf_plot}
\end{figure}
Figure \ref{cdf_plot} shows the cumulative distributions of the line-of-sight velocity ($v$) and centre-to-limb angle ($\mu = \cos\theta$). The cumulative distributions of $v$ show how the rotational velocities are distributed across latitude bands and that the range of velocities decreases from 2 km/s at 0$\degree$ latitude to 1 km/s at 60$\degree$ latitude. The cumulative distributions of $\mu$ show the projection effect as we move from the centre of the solar disk.




\bsp	
\label{lastpage}
\end{document}